\documentclass[11pt]{article}
\pdfoutput=1
\usepackage[letterpaper, margin=1in]{geometry}
\usepackage{cite}
\usepackage[font=small,labelfont=bf]{caption}
\usepackage[xcolor,leftbars]{changebar}
\usepackage{fancyhdr}

\usepackage{graphicx,amsmath,amsfonts,amssymb,slashed}
\usepackage{caption}
\usepackage{subcaption}
\usepackage{numprint}
\usepackage{enumitem}
 \usepackage{mathtools}        
\usepackage{color}
\usepackage[bookmarks=false]{hyperref}
\usepackage{xcolor}

\usepackage[force]{feynmp-auto}
\usepackage{tikz}
\usepackage[compat=1.0.0]{tikz-feynman}
\usetikzlibrary{positioning}

\newcommand{\beq}{\begin{equation}}
\newcommand{\eeq}{\end{equation}}
\newcommand{\ie}{{\it i.e. }}
\newcommand{\eg}{{\it e.g. }}

\definecolor{UOgreen}{HTML}{007030}

\hypersetup{ 
colorlinks, 
linkcolor={UOgreen}, 
citecolor={UOgreen}, 
urlcolor= [rgb]{0.75 0.05 0.05} 
}

\graphicspath{{figs/}}

\title{2024 TASI Lectures: A Dark Matter Primer}
\author{Tien-Tien Yu\\
{\small \color{UOgreen}\texttt{tientien@uoregon.edu}, }\\
{\small Institute for Fundamental Science and Department of Physics,}\\
{\small University of Oregon, Eugene, OR 97403}
}
\date{}

\begin{document}

\maketitle
\pagestyle{headings}

\abstract{
These notes are based on a sequence of 4 lectures delivered at the 2024 Theoretical Advanced Study Institute (TASI) and at the Università degli Studi di Padova. They are intended for graduate students at the early stages of their study of dark matter with some prior exposure to cosmology and quantum field theory. The primary aim is to offer an accessible introduction to dark matter and to lay the groundwork for exploring its phenomenology. These lectures are not intended to serve as a comprehensive review; one can find a recent and more in-depth treatment of dark matter phenomenology in Ref.~\cite{Cirelli:2024ssz}.
We begin by motivating the study of dark matter through a discussion of the empirical evidence and the constraints it places on dark matter properties. This is followed by an overview of several canonical mechanisms for the production of cosmological dark matter, and a concluding section that surveys current experimental and observational efforts to detect it with a focus on direct detection. 
These notes have benefited from the extensive body of literature on dark matter, particularly the discussions presented in Refs.~\cite{Lin:2019uvt,doi:10.1142/q0001,Murayama:2007ek,TaitSSI2020:1}.
%
\section{What is Dark Matter?}\label{sec:what}
\subsection{Evidence for Dark Matter}\label{sec:evidence}
Dark matter (DM) is a non-relativistic fluid\footnote{A fluid can be described by the equation of state $P=w\rho$, where $P$ is the pressure and $\rho$ is the energy density. For relativistic species $w=1/3$ while for non-relativistic matter $w=0$.} present in the cosmos, and its existence is necessary to explain the dynamics and evolution of our Universe. We know this through various forms of observational evidence, which span across many scales. To date, all experimental evidence for the existence of DM comes from astrophysical observation. These observations also provide clues to the properties of the DM. 

At the smallest scales, we have evidence from {\bf galactic rotation curves}. According to Newton's 2nd Law, if most of a galaxy’s mass were concentrated in the visible stars and gas near the center, then the orbital velocity \( v(r) \) of stars at a distance \( r \) from the center should decrease as:

\begin{equation}
    v(r) \approx \sqrt{\frac{G_NM(r)}{r}}
\end{equation}
where \( M(r) \) is the mass enclosed within radius \( r \) and $G_N=6.674\times 10^{-11}$m$^3$kg$^{-1}$s$^{-2}$ is Newton's Gravitational constant. However, measurements of the rotation curves of spiral galaxies using Doppler shifts in emission lines from stars and gas, show that the rotational velocities remain nearly constant at large radii, rather than falling off as expected~\cite{BinneyTremaine2008,Bosma1978,RubinFord1970,1989A&A...223...47B,Sofue:2000jx,Salucci:2018hqu,10.1093/mnras/278.1.27}. If there is additional unseen mass distributed in a large halo around the galaxy, then \( M(r) \) continues to increase at large distances, counteracting the expected velocity drop. This halo of mass is what we call DM. 
\footnote{While DM is the leading explanation, some alternative theories have been proposed, including Modified Newtonian Dynamics (MOND), in which one modifies gravity at low accelerations to fit observation. However, MOND does not naturally explain key astrophysical and cosmological observations, so we will not discuss it further. The interested reader can find more information in Refs.~\cite{Famaey:2011kh,Famaey:2025rma}.} 

Moving up in size, we can look at {\bf galaxy cluster mergers} to further support the need for DM and to glean more information about its nature. Specifically, galaxy cluster mergers demonstrate that the majority of mass in galaxy clusters does not behave like ordinary (baryonic) matter. Using X-ray telescopes (\eg Chandra), we observe that most of the baryonic matter (hot gas) is concentrated in the center of the system as the gas experiences strong interactions. The total mass distribution is mapped through gravitational lensing. These lensing maps reveal that the bulk of the mass is not aligned with the hot gas region, but rather in two separate regions aligned with the galaxies. This suggests that a large fraction of the cluster's mass is (nearly) collisionless and interacts gravitationally, but not via electromagnetic or nuclear forces. In Fig.~\ref{fig:bulletcluster}, we see one such example of a galaxy cluster merger known as the Bullet Cluster. The Bullet Cluster reveals a striking separation between the visible matter (hot gas, detected via X-rays and shown in red and yellow) and the mass distribution (mapped through gravitational lensing and shown in blue). This clear offset demonstrates that most of the cluster's mass, determined by the gravitational lensing, is not comprised of baryonic gas. Note that modified gravity theories struggle to explain the observed mass distribution.
\begin{figure}
    \centering
    \includegraphics[width=0.95\linewidth]{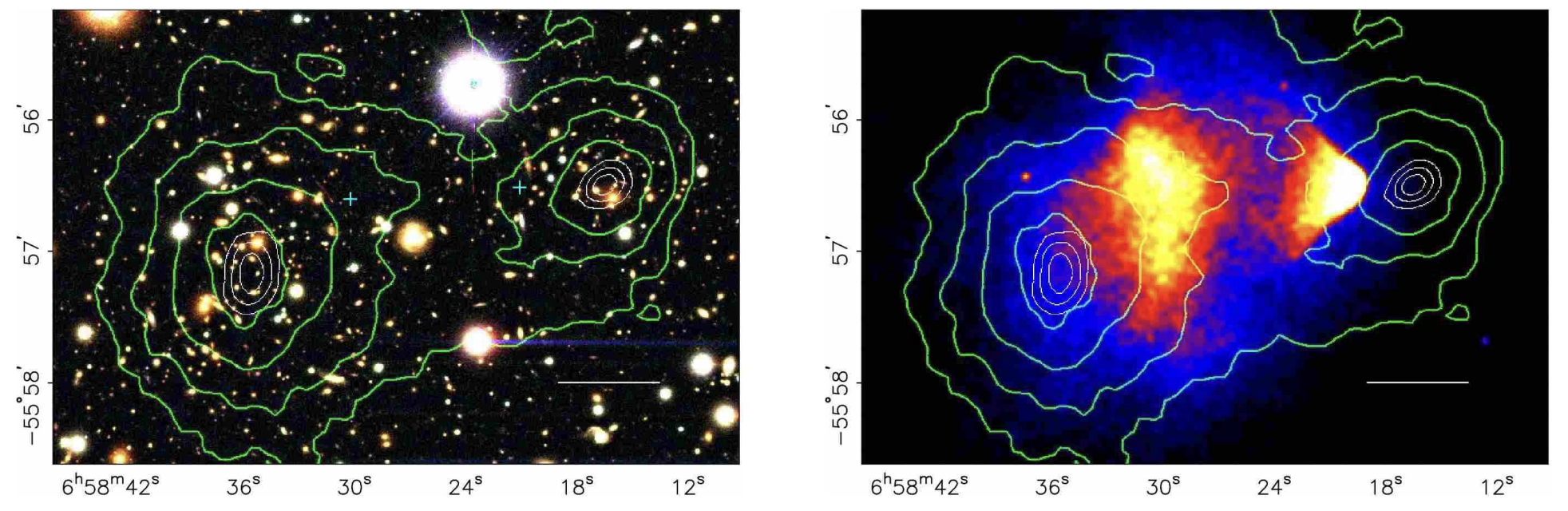}
    \caption{Shown in both panels are green contours of the weak-lensing reconstruction of the mass distribution. The ({\bf left}) shows the optical image of the background galaxies while the ({\bf right}) panel shows the X-ray imaging of the baryonic matter of the two galaxy clusters. Figure reproduced with permission from Ref.~\cite{Clowe:2006eq}
}
    \label{fig:bulletcluster}
\end{figure}

At even larger scales, we can study {\bf structure formation}. Observations of large-scale structure from galaxy surveys like the Sloan Digital Sky Survey (SDSS) closely match predictions from cosmological simulations such as the Millennium Simulation, which assumes a universe dominated by cold DM (CDM) (see \eg Ref.~\cite{Springel2006}).
 The distribution of galaxies in SDSS follows a cosmic web structure of filaments, voids, and clusters, consistent with the hierarchical growth of structure expected in a $\Lambda$CDM cosmology~\cite{Springel2005,Springel2006}. The power spectrum of galaxy clustering and Baryon Acoustic Oscillation (BAO) features observed in SDSS provide independent confirmation that DM played a crucial role in amplifying early density fluctuations into the structures we see today~\cite{SDSS:2005xqv}, while simulations without DM fail to reproduce these structures~\cite{Davis:1985rj}.

And at the largest of observable scales, we have the {\bf Cosmic Microwave Background (CMB)}, which provides some of the most compelling evidence for the existence of DM. The CMB power spectrum exhibits distinct acoustic peaks, which arise due to oscillations in the early Universe’s photon-baryon plasma. The relative heights and positions of these peaks can only be accurately explained if a significant fraction of the Universe's matter is in a form that does not interact electromagnetically, \ie DM. Specifically, DM influences the first peak's height, which is sensitive to the total matter density, and determines the third peak’s amplitude. The overall shape of the power spectrum constrains the cosmic matter density, $\Omega_m$, and demonstrates that ordinary baryonic matter alone is insufficient to account for the observed structure~\cite{Planck:2018vyg,WMAP:2003elm}. 
Fig.~\ref{fig:CMB} presents data from the Planck satellite alongside the predicted power spectrum for a universe where DM accounts for approximately 25\% of the total energy density.
\begin{figure}
\centering
\includegraphics[width=0.6\linewidth]{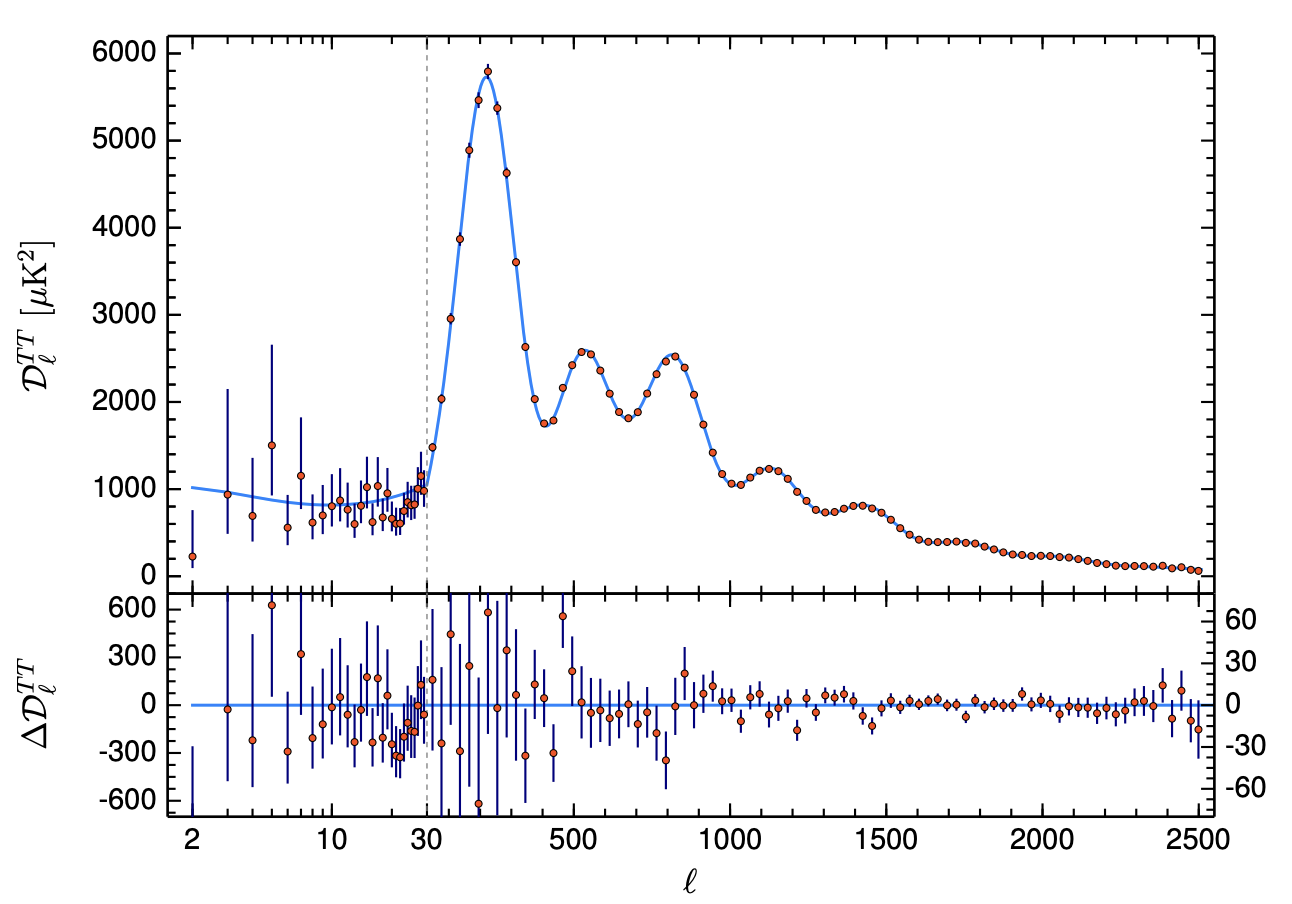}
\caption{{\it Planck} 2018 temperature power spectrum. The light blue curve is the best fit theoretical $\Lambda$CDM spectrum, which gives us the relative abundance of baryonic matter, dark matter, and dark energy in the cosmic energy budget. Figure reproduced with permission from Ref.~\cite{Planck:2018vyg}}
\label{fig:CMB}
\end{figure}

All direct observational evidence for DM relies on its {\it gravitational} influence, implying it must have mass and behave like pressureless matter (hence the name). Beyond the fact that DM is some massive object, what other properties must DM satisfy? 

\subsection{Properties of Dark Matter}\label{sec:properties}
{\bf Dark matter is ``dark": }
We have already established that DM must behave like matter, which explains the second word in the nomenclature. We also require that DM is ``dark" with respect to SM interactions, giving rise to the first word. What does this mean? This means that DM is not luminous in galaxies or galaxy clusters that we observe today. Furthermore, observations of the matter power spectrum and the CMB require that the matter component of the Universe to have primarily gravitational interactions. Let's unpack this last point in more detail. 

To proceed with the discussion, we will introduce a key quantity known as the {\it Matter Power Spectrum}, denoted as ${\cal P}(k)$ (for a more detailed discussion, see Ref.~\cite{Dodelson:2003ft}). The matter power spectrum quantifies the statistical distribution of density fluctuations in the Universe, characterizing the variance of the density contrast $\delta(\mathbf{k})$ in Fourier space. It is a function of the co-moving wavenumber $k$, which corresponds to a characteristic length scale $2\pi/k$. Larger values of ${\cal P}(k)$ indicate greater variance in density fluctuations at the corresponding scale, implying stronger clustering of matter. On large scales, the growth of structure is governed by the interplay between gravitational attraction and cosmic expansion, with density fluctuations evolving according to linear theory. More commonly, analyses use the dimensionless power spectrum, which provides a normalized measure of these fluctuations:
\beq
\Delta^2(k)\equiv 4\pi\left(\frac{k}{2\pi}\right)^3{\cal P}(k)\, .
\eeq
Note that $\Delta^2(k) \sim 1$ corresponds to ${\cal O}(1)$ density fluctuations, meaning that the overdensity is comparable to the average density and signals the onset of nonlinear structure formation. The density fluctuations provide information about gravitational collapse, where large initial perturbations grow and eventually collapse into bound structures such as halos. The thermal history of DM in the early Universe plays a crucial role in determining the cosmological clustering of DM on small scales. If DM is a cold thermal relic that becomes non-relativistic soon after decoupling from the SM plasma, it can condense into low-mass structures. However, if DM remains relativistic for longer (\ie retains significant kinetic energy), it will free-stream through low-mass density perturbations and suppress structure formation on small scales. This suppression continues until the DM velocity decreases due to momentum redshifting from cosmic expansion or kinetic decoupling from the thermal bath.

The matter power spectrum $\mathcal{P}(k)$, and consequently $\Delta^2(k)$, evolves with redshift. Observational constraints on $\mathcal{P}(k)$ span a range of scales ($k \sim 10^{-3} - 10$ Mpc$^{-1}$), with measurements derived from the Cosmic Microwave Background (CMB, $z \sim 1100$)~\cite{Dodelson:2003ft}, the Lyman-$\alpha$ forest ($z \sim 2-6$)~\cite{Viel:2005eg, Irsic:2017yje}, and large-scale structure surveys ($z \sim 0-3$)~\cite{eBOSS:2020yzd}. A given Fourier mode $k$ enters the horizon when its comoving scale satisfies $k = aH$, marking the transition from superhorizon to subhorizon evolution. 
Earlier evolution, which corresponds to higher redshift, leads to larger density perturbations. The power spectrum at small scales (large $k$) provides insight into the microphysics of DM, particularly whether it experienced interactions with SM particles at early times, which would suppress power at large $k$ through collisional damping or free-streaming effects. If DM had significant interactions with SM particles in the early Universe, this would damp small-scale structure formation. This effect can be seen in Fig.~\ref{fig:powerspectrum}, where the blue line labeled {``interacting DM"} exhibits suppressed power at large $k$. Observations from the CMB, Lyman-$\alpha$ forest, and large-scale structure surveys (\eg SDSS) place strong constraints on DM interactions with SM particles, which indicates that any significant interactions must have ceased well before recombination, likely at much earlier epochs.

\begin{figure}
    \centering
    \includegraphics[width=0.65\linewidth]{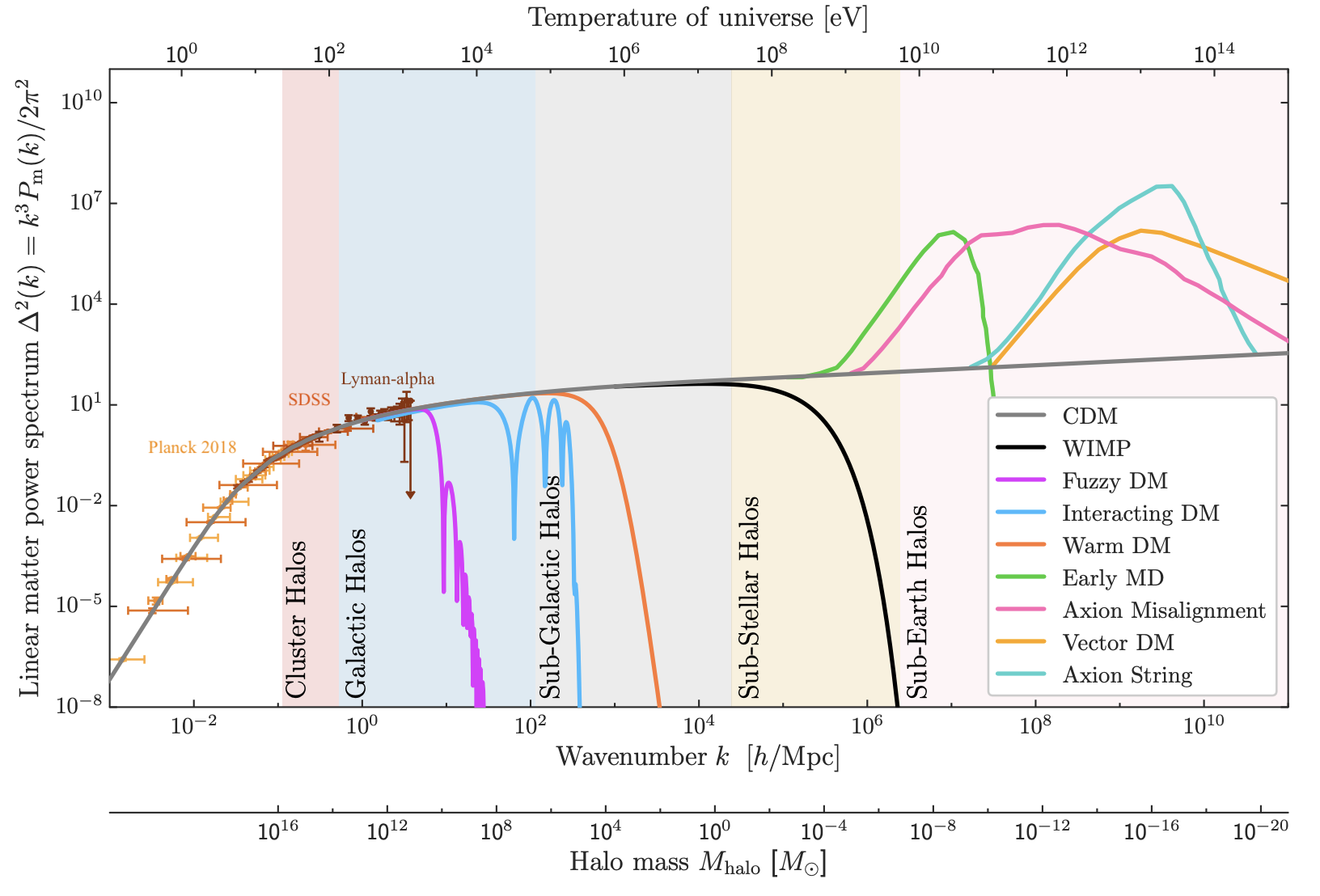}
    \caption{The dimensionless linear matter power spectrum extrapolated to $z = 0$. The shape of the linear matter power spectrum (colored lines) can inform us about the fundamental nature of DM such as its mass and interactions. The shaded regions indicate the size of the DM halos probed at that scale. The data -- from Planck 2018, DES Y1, SDSS DR7 LRG, and eBOSS DR14 -- is compiled from Ref.~\cite{Chabanier:2019eai}. Figure adapted from Ref.~\cite{Bechtol:2022koa}. }
    \label{fig:powerspectrum}
\end{figure}

{\bf Dark matter is non-relativistic, \ie cold}: There are a few ways to come to the conclusion that DM must be non-relativistic in the early Universe. Again, we can arrive at this conclusion through $\Delta^2(k)$. If DM is relativistic, then the perturbations within a horizon can become washed out due to the motion of DM. Therefore, there would be a relative suppression in $\Delta^2(k)$ for modes which enter the horizon when DM is still relativistic. In Fig.~\ref{fig:powerspectrum}, we can see this in the dark orange line labeled ``Warm DM." If the DM was in thermal equilibrium with similar temperature to the photons, the bounds from Lyman-$\alpha$ measurements require that $m_\chi\gtrsim$ keV. 

We can also come to the conclusion that DM is non-relativistic by looking at galactic rotation curves. First, note that the galactic escape velocity is about 544 km/s~\cite{ParticleDataGroup:2018ovx}; any particle going faster than this would not be bound to our galaxy, and therefore would not be around to explain the behavior of galactic rotations curves. To expand on our earlier discussion, we start with Newton's second law, which relates the circular velocity of a test mass orbiting a central mass to the enclosed source mass, $v_c(r)=\sqrt{G_NM(r)/r}$. 
In the case of galaxies, the source mass $M$ corresponds to the total mass enclosed within a given radius, including contributions from both visible matter and any unseen mass components. 

Far away from the galaxy, the mass of the galaxy $M(r)=$ constant, and so we have $v_c(r)\propto 1/\sqrt{r}$, as seen in the left panel of Fig.~\ref{fig:rotation_ex}. Inside the galaxy, the mass of the (spherical) galaxy is given by $M(r)=4\pi\int_0^r dr' r'^2 \rho(r')$. If we assume that $\rho(r)=$ constant, then we find that $v_c(r)\propto r$, as seen in the right panel of Fig.~\ref{fig:rotation_ex}. 
\begin{figure}
    \centering
    \includegraphics[width=0.45\linewidth]{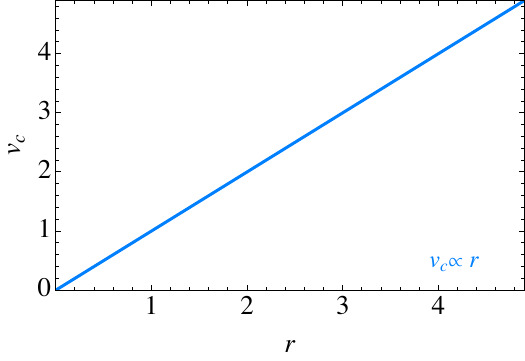}
\includegraphics[width=0.45\linewidth]{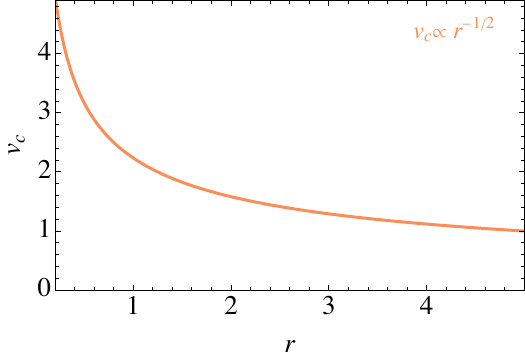}
    \caption{Examples of rotations curves for ({\bf left}) a test mass inside a galaxy of constant density and ({\bf left}) a test mass outside of the galaxy. 
    }
    \label{fig:rotation_ex}
\end{figure}

Now let's look at observation. We can obtain the rotation curves of various galaxies through several means, 
including the 21cm emission from atomic hydrogen, 
optical emission lines from hotter gas,
and optical absorption lines from the stellar component~\cite{BinneyTremaine2008,Bosma1978,RubinFord1970,1989A&A...223...47B,Sofue:2000jx,Salucci:2018hqu,10.1093/mnras/278.1.27}. 
Fig.~\ref{fig:M33} shows the predictions the rotation curves from the stellar and gas components of the galaxy compared to the observations. By adding in a DM component, labeled NFW halo, the prediction more closely aligns with the observational data. 
There are two things to note: 
\begin{enumerate}
    \item The data does not fall off like $1/\sqrt{r}$ despite the surface density, \ie the visible disk, going to zero.
    \item The data is roughly flat at large radius.
\end{enumerate}
The lessons we can learn from these two things are that
\begin{enumerate}
    \item the matter distribution extends well-past the visible disk.
    \item the mass distribution of the galaxy goes roughly like $M(r)\propto r$, which implies that the mass density $\rho(r)\propto r^{-2}$ if we assume a spherically-symmetric galaxy.
\end{enumerate}
Now let's apply this information to our Milky Way galaxy. From stellar kinematics, we can get the total mass of the Milky Way halo and the DM density, $M_{halo}\sim 10^{12}M_{\odot},~\rho_{\rm DM}\sim 0.4$ GeV/cm$^3$. Taking a spherically-symmetric galaxy, we can calculate the halo radius,
\beq
M_{\rm halo}\sim 4\pi\int_0^{R_{\rm halo}}dr r^2\rho_{\rm DM}(r)\implies R_{\rm halo}\sim 100{~\rm kpc}\, .
\eeq
To get the average velocity of the DM in our Milky Way halo, we apply the virial theorem,
\beq
\langle v_{\rm DM} \rangle \sim \sqrt{\frac{G_NM_{\rm halo}}{R_{\rm halo}}}\sim 200{~\rm km/s}\, .
\eeq
This is well within the non-relativistic regime.

\begin{figure}
    \centering
\includegraphics[width=0.65\linewidth]{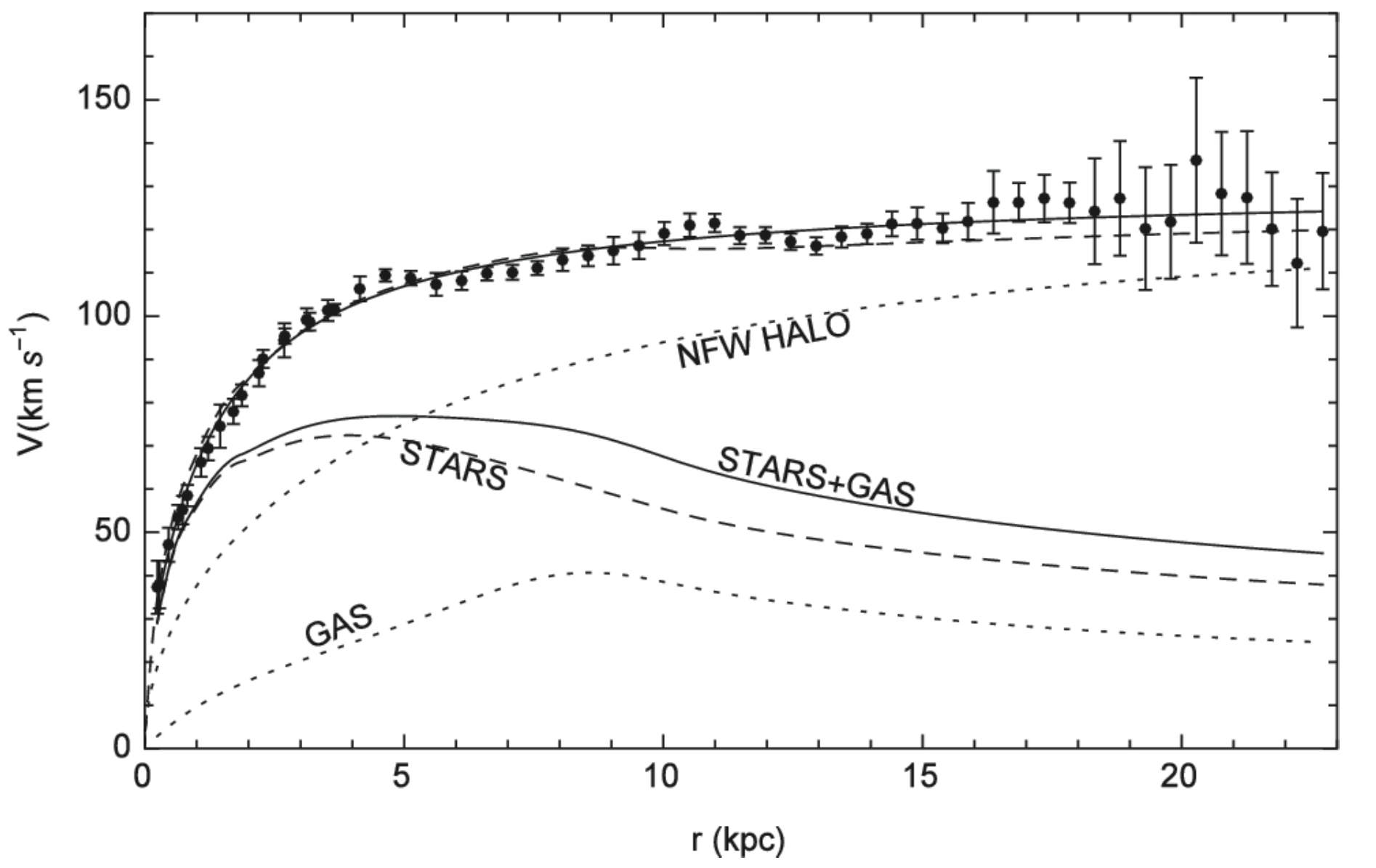}
\caption{Rotation curve of M33 ({\bf black} dots with error bars) compared to the stellar and gas contributions. The dotted curve labeled NFW halo is the contribution from DM. Figure reproduced with permission from Ref.~\cite{LopezFune:2016uun}.}
    \label{fig:M33}
\end{figure}

{\bf Dark Matter is collisionless on large scales:} Again, we can glean this property of DM from $\Delta^2(k)$. Non-gravitational interactions within the dark sector would lead to modifications in $\Delta^2(k).$ The main observational constraints come from galaxy and galaxy cluster scales (see Ref.~\cite{Tulin:2017ara} for an extensive review). Suppose we have two halos with sizes $L\sim $Mpc and masses $M_{\rm halo}\sim 5\times 10^{14} M_\odot$. The probability for DM particle $\chi$ to scatter with another DM particle
on the typical scale of the system is given by $P_{\chi\chi\to\chi\chi}\sim n_\chi\sigma_{\chi\chi\to\chi\chi} L$ where $n_\chi\sim M_{\rm halo}/m_\chi \left(4/3\pi L^3\right)^{-1}$. Requiring that $P_{\chi\chi\to\chi\chi}<1$ sets that $\sigma_{\chi\chi\to\chi\chi}/m_\chi\lesssim 10$ cm${^2}$/g. A more careful modeling of the Bullet Cluster gives $\sigma_{\chi\chi\to\chi\chi}/m_\chi\lesssim 1.25{~\rm cm}{^2}$/g~\cite{Randall:2008ppe}.\footnote{This bound changes if there is any non-trivial velocity-dependence in the DM scattering cross-section~\cite{Sagunski:2020spe}.}

{\bf Dark Matter is stable on cosmological scales:} since we have DM around today, we know that its lifetime must  be at least longer than the age of the Universe. Given that the age of the Universe is $t_{\rm Universe}\simeq 13.8{~\rm Gyr}= 4.35\times 10^{17}$ s, we can determine that $\Gamma_\chi\lesssim 3\times 10^{-18}$ s$^{-1}$.~\footnote{Indirect DM searches, which look at the annihilation of DM into ``cosmic messengers", such as photons, neutrinos, and cosmic rays, can constrain lifetimes much greater than $10^8\times$ (age of the Universe)~\cite{Cooley:2022ufh}.} This tells us that there must be some mechanism that prevents DM from decaying. One possibility is that there is some symmetry (or at least approximately) that prevents DM from decaying, such as a $\mathbb{Z}_2$ symmetry that takes $\chi\to -\chi$ but SM$\to$SM.

{\bf Dark Matter forms halos:} we can also learn some general, model-independent properties about DM based on the fact that DM forms halos. 
For bosons, the limit can be obtained by invoking the uncertainty principle, $\Delta x\Delta p\geq 1$. Since we require that the DM be contained within the halo, $\Delta x=R_{\rm halo}$, where $R_{\rm halo}$ is the virial radius of the DM halo, while $\Delta p=m_\chi \Delta v_\chi$, with $\Delta v_\chi$ the velocity dispersion of the DM halo. Taking typical values from a dwarf galaxy, with $R_{\rm halo}\sim$ kpc and $\Delta v_\chi\sim 10$ km/s, we get $m_\chi\gtrsim 10^{-21}$ eV. More precise modeling and analysis pushes the lower limit up to $2.2\times 10^{-21}$ eV~\cite{Zimmermann:2024xvd}.

What about for fermions? Here, we are subject to Fermi-Dirac statistics at the Pauli exclusion principle such that $f(\vec x,\vec p)\leq g\hbar^3$, where $g=$ number of spin and flavor states. The local DM number density is then given by
\beq
n(\vec x)=\int\frac{d^3p}{(2\pi)^3}f(\vec x,\vec p)\leq \frac{g}{8\pi^3}\frac{4\pi}{3}p^3_{\rm max}\equiv n_{\rm max}\, ,
\eeq
where $p_{\rm max}\leq m_\chi v_{\rm esc}$ is the maximum possible momentum.  

Now let us consider a virialized halo, where $v_{\rm esc}\simeq \sqrt{\frac{2G_NM_{\rm vir}}{R_{\rm vir}}}$. The average density of such a halo is given by $n_{\rm avg}\sim \frac{3M_{\rm vir}}{4\pi R_{\rm vir}^3 m_\chi}$. Requiring that $n_{\rm avg}\leq n_{\rm max}$ gives us
\beq
\frac{3M_{\rm vir}}{4\pi R_{\rm vir}^2m_\chi}\leq \frac{g}{8\pi^3}\frac{4\pi}{3}m_\chi^3\left(\frac{2G_NM_{\rm vir}}{R_{\rm vir}}\right)^{3/2} \implies m_\chi^4\geq \frac{5}{g}M_{\rm vir}^{-1/2}R_{\rm vir}^{-4/2}G_N^{-3/2}\, .
\eeq
For the Milky Way, $M_{\rm vir}\sim 10^{12}M_\odot,~R_{\rm vir}\sim 0.3$ Mpc $\implies m_\chi\geq 5$ eV. Plugging in the values for the Fornax dwarf galaxy, we can obtain an even stronger abound of $m_\chi\geq 70$ eV. A recent analysis refines these bounds to roughly $m_\chi\geq 180$ eV~\cite{Alvey:2020xsk}. This is known as the ``Tremaine-Gunn" bound. 

What about an {\bf upper bound for DM mass}? Suppose we have a DM point particle with mass $m_\chi$. The Schwarzschild radius of such an object is given by $R_S=2G_Nm_\chi$. The uncertainty principle implies that the particle cannot be localized to better than its Compton wavelength $R_c=1/m_\chi$. If $m_\chi$ is small, then we satisfy $R_c\gg R_S$. What if $m_\chi$ was close to the Planck mass? More specifically, let's consider the scenario in which $m_\chi\geq G_N^{-1/2}\sim 10^{19}$ GeV$\sim M_{\rm Pl}$. In this case, $R_c<R_S$ and the particle is localized within the Schwarzschild radius, resulting in a finite-sized black hole. Therefore, we can conclude that $m_\chi < M_{\rm Pl}$ for a point particle. 

However, DM does not necessarily have to be a point particle, but could instead be some composite or macroscopic object. There are also limits to the mass of DM in these scenarios.
Massive (greater than $M_\odot$) objects could affect the stability of bound systems. Therefore, $m_\chi > 10^3M_\odot~(10^{60}{~\rm eV})$ would disrupt globular clusters~\cite{Goerdt:2006hp}. A generic upper mass limit comes from searches for MAssive Compact Halo Objects (MACHOs), which result in gravitational lensing signatures. Notably, the EROS and MACHO collaborations have combined their data to set stringent constraints on the abundance of MACHOs in the Galactic halo. Their findings indicate that MACHOs with masses in the range of $10^{-6}M_\odot$ to $10^2 M_\odot$ cannot constitute more than approximately 15\% of the total halo mass~\cite{1997A&A...324L..69R, 2007A&A...469..387T, Blaineau:2022}.

As discussed before, DM cannot be hot (relativistic in the early Universe). Another argument for this is that hot DM destroys clusters smaller than $\sim10^{15} M_\odot$~\cite{White:1983fcs}. A consequence of this statement is that neutrinos cannot be (all of) DM! We will revisit this argument in Section~\ref{sec:creating}.

In Fig.~\ref{fig:mass_landscape}, we summarize the potential mass parameter space for DM. The different colors demarcate possible production mechanisms for DM, which we will discuss in the next section. 
\begin{figure}
    \centering
\includegraphics[width=0.75\linewidth]{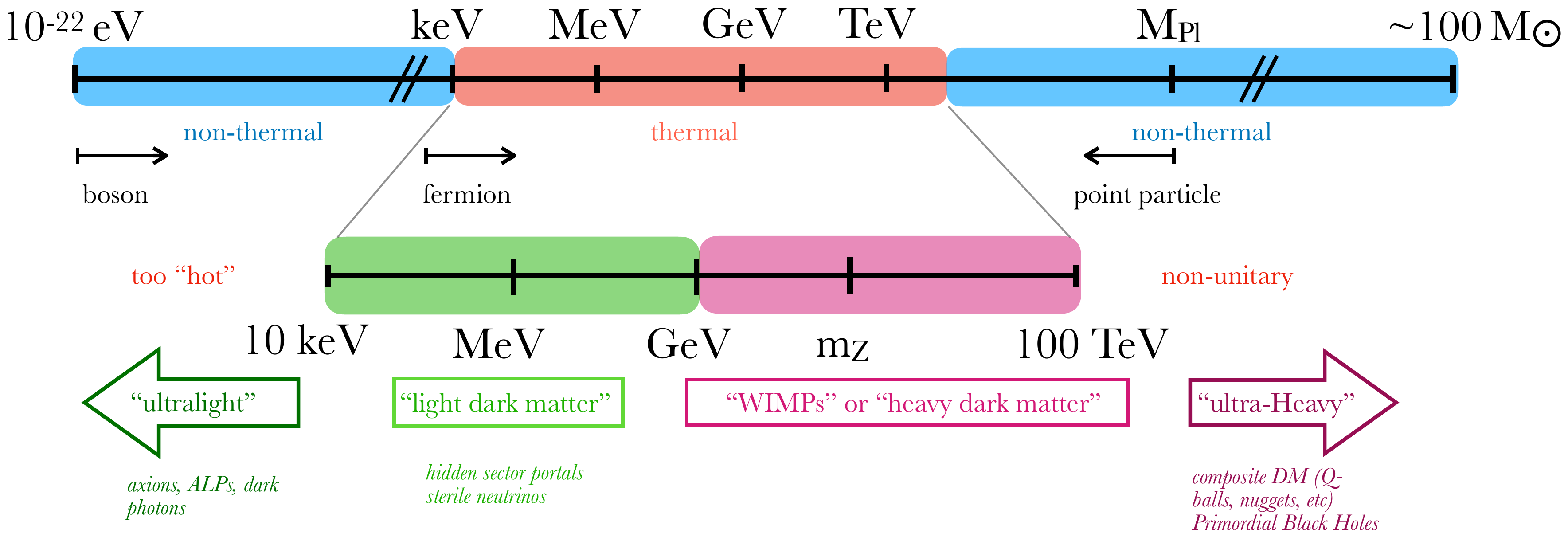}
    \caption{Dark Matter mass landscape as determined by the observational evidences described in Section~\ref{sec:evidence}. The thermal ({\bf red}) and non-thermal ({\bf blue}) regions denote categories of cosmological production, which are discussed in Section~\ref{sec:properties}. The {\bf bottom row} provides four categories of DM candidates: ultralight ($m_\chi\lesssim 1$ eV), light (1 eV$\lesssim m_\chi\lesssim 1$ GeV), heavy (1 GeV$\lesssim m_\chi\lesssim 100$ TeV), and ultra-heavy (100 TeV$\lesssim m_\chi$)~\cite{Cooley:2022ufh}.}
    \label{fig:mass_landscape}
\end{figure}
\section{Creating Dark Matter}\label{sec:creating}
In the previous section, we saw that there is a wide mass range for candidates of DM. In this section, we will discuss the creation of DM and how this provides us a rubric in prioritizing our searches for DM. Specifically, one can require that the most compelling theories are the ones with viable cosmological histories. 

The relic abundance of DM is measured at~\cite{Planck:2018vyg}
\begin{equation}
    \Omega_{\chi} h^2=0.120\pm 0.001\, ,
\end{equation}
where $H_0=h\cdot 100$ km/s/Mpc is the rate of expansion today, with $h=0.68$.\footnote{Early and late-time measurements give discrepant values for $h$, see Ref.~\cite{DiValentino:2021izs} for a recent review} Here, $\Omega_{\chi}=\rho_{\chi}/\rho_c$, where $\rho_c\equiv3H_0^2/(8\pi G)\simeq 1.05\times 10^{-5}h^2$ GeV cm$^{-3}$ is the critical density of the Universe. Given the value of $h$, we can read off that $\Omega_{\chi}=0.264\pm0.003$, meaning that DM constitutes roughly 26\% of the matter-energy content of the Universe today. 

For much of its early history, most constituents of the Universe were in thermal equilibrium, forming what is called a ``thermal bath." Departures from equilibrium lead to important relics, such as the light elements, $\nu$-backgrounds, and net baryon number.
The discussion about creating DM can be framed in terms of this thermal bath. Specifically, we can ask if the DM is produced, as well, from this thermal bath. 
If so, we can describe the evolution of the DM abundance through its Boltzmann equation. 

In full thermal equilibrium, the number density \( n(t) \) satisfies the condition:
\begin{align}
    \frac{d}{dt}\left[n(t)a(t)^3\right] = 0 \rightarrow 
    \dot{n}(t) + 3H(t)n(t) = 0,
\end{align}
where $a(t)$ is the scale factor that governs the expansion of the Universe and $H\equiv\dot a/a$.

Departures from equilibrium occur when number-changing processes become relevant. The general form of the Boltzmann equation is:
\begin{equation}
    \dot{n} + 3Hn = -\Gamma(n - n_{\rm eq}) - \langle\sigma v\rangle (n^2 - n_{\rm eq}^2),
\end{equation}
where \( \Gamma \) is a decay rate, \( \langle\sigma v\rangle \) is the thermally averaged interaction cross-section, and \( n_{\rm eq} \) is the equilibrium number density. Note that these $1\to 2$ and $2\to2$ processes will a priori dominate over processes with larger number of initial particles, such as $3\to2$, but the latter could contribute if the decay or annihilation channels are not open. 

This equation is insightful in different limiting cases:
\begin{itemize}
    \item If \( \Gamma \) or \( \langle\sigma v\rangle \) is large compared to \( H \), the DM number density \( n \) remains close to its equilibrium value \( n_{\rm eq} \).
    \item If \( H \) is much larger than the interaction rates, \( n \) effectively ``freezes out" and evolves primarily due to the Universe's expansion.
\end{itemize}

As a specific example, 
consider the number-changing process \( \chi_1\chi_2 \to \chi_3\chi_4 \). The evolution of the number density \( n_1 \) is given by:
\begin{equation}
\label{eq:n1evol}
    a^{-3} \frac{d(n_1 a^3)}{dt} = \langle\sigma v\rangle n_1^0 n_2^0 \left(\frac{n_3 n_4}{n_3^0 n_4^0} - \frac{n_1 n_2}{n_1^0 n_2^0} \right),
\end{equation}
where \( \langle\sigma v\rangle = \langle\sigma(\chi_1\chi_2 \to \chi_3\chi_4) v\rangle \), and the superscript \( 0 \) denotes thermal equilibrium values.

If we assume \( n_3 = n_3^0 \), \( n_4 = n_4^0 \), and define \( n_1 = n_2 \equiv n_\chi \), the equation simplifies to:
\begin{equation}
    a^{-3} \frac{d(n_\chi a^3)}{dt} = \langle\sigma v\rangle \left[ (n_\chi^0)^2 - n_\chi^2 \right].
\end{equation}

A useful quantity to introduce is the comoving number density, \( Y \equiv n/s \), where \( s = (P+\rho)/T=(1+w)\rho/T \) is the entropy density. Defining \( x \equiv m_\chi / T \) and using entropy conservation (\( sa^3 = \) constant for particles in equilibrium), we obtain:
\begin{equation}
\label{eq:Yevol}
    \frac{dY}{dx} = x \frac{s(m_\chi)}{H(m_\chi)} \langle\sigma v\rangle \left[ (Y^0)^2 - Y^2 \right].
\end{equation}

From this, we can glean a few key takeaways:
\begin{itemize}
    \item We have an expression for \( Y \) as the Universe cools.
    \item \( Y \) represents the DM number density rescaled by entropy density, removing the explicit dependence on cosmic expansion (\( H \) no longer appears in Eq.~\ref{eq:Yevol}).
    \item Any change in \( Y \) arises from DM interactions with particles in thermal equilibrium with the photon bath, $Y^0$.
    \item The evolution of \( Y \) is governed by the velocity-averaged annihilation cross-section \( \langle\sigma v\rangle \).
\end{itemize}

Finally, from \( Y \), we can compute the DM relic abundance as a fraction of the critical density:
\begin{equation}
    \Omega_\chi = \frac{m_\chi s Y}{\rho_c}=0.264\, .
\end{equation}
We can gain some intuition by plugging in the values for the entropy density contained in photons $s_{\gamma,\rm today}=1541.9{~\rm cm}^{-3}=1.1\times 10^{-11}$ eV$^3$ and $\rho_c=4.9\times 10^{-6}$ GeV/cm$^3$=$3.7\times 10^{-11}$ eV$^4$ to get 
\begin{equation}
\label{eq:Ytoday}
Y_{\rm today}\simeq \frac{\rm{eV}}{m_\chi}\, . \end{equation} 
With the concept of thermal equilibrium in hand, we can now classify DM candidates in terms of answers to the following question: 
\begin{center}
\fbox{
\large{\bf Was the DM ever in thermal equilibrium with the Standard Model?}
}
\end{center}
   
\subsection{Yes: Thermal relics and freeze-out}\label{sec:freezeout}
For our first scenario, let's suppose the answer to the above question is {\it yes}. These candidates are what are known as thermal relics. From the discussion about structure formation and limits from Ly-$\alpha$, we saw that there is a minimum mass of $m_\chi\gtrsim$ keV. Thermal relics interact with the SM through reactions that keep them in ``thermal equilibrium" in the early Universe. As the Universe cools and expands, the number density of the DM particles decreases. At some point, the rate of the reactions, $\Gamma$, is smaller than the expansion rate of the Universe and the particle goes through ``thermal decoupling." At this point, the particle is no longer in thermal equilibrium and has ``frozen-out." 

\subsubsection{Neutrinos}\label{sec:neutrinos}

To begin, let's take a look at the SM neutrinos. A priori, the SM neutrinos satisfy many of the essential requirements of a DM candidate: they are matter, they are stable (or at least very long-lived), and they are ``dark" (\ie do not experience electromagnetic or strong interactions). 

The relevant reaction is $\Gamma:e^+e^-\to \nu\bar\nu$, which proceeds through the weak interaction. 
From dimensional analysis, we can determine that the cross-section parametrically goes like $\langle\sigma v\rangle\sim G_F^2 T^2$, where $G_F\simeq 1.166\times 10^{-5}$ GeV$^{-2}$ is the Fermi coupling constant, and the rate is given by $\Gamma\simeq n_\nu\langle\sigma v\rangle$.
To determine the temperature of freezeout, $T_\nu$, we want to solve for when $\Gamma(T_\nu)=H(T_\nu)$. Below this temperature, the rate of the neutrino interactions is slower than the expansion of the Universe, so the number of neutrinos freezes-out. For a relativistic species, we know that the number density $n_{\rm rel}\sim T^3$ for $m\ll T$~\cite{Kolb:1990vq}. Hubble is defined as $H^2\equiv \frac{8\pi G_N}{3}\rho$, where $G_N$ is the gravitational constant and $\rho$ is the energy density, which gives $H\simeq 1.66\sqrt{g_*(T)}\frac{T^2}{M_{\rm Pl}}$ in a radiation-dominated Universe with $g_*(T)$ is the effective number of relativistic degrees of freedom. 
Combining these scalings, we have that $\Gamma\sim G_F^2 T^5$ and $H\sim T^2/M_{\rm Pl}$, from which we can read off that $T_\nu\simeq (G_F^2 M_{\rm Pl})^{-1/3}\simeq 1$ MeV $\gg m_\nu$. Therefore, neutrinos freeze-out while they are relativistic and constitute ``hot" thermal relic DM. Note, importantly, that $T_\nu\ll m_{W,Z}$ so the 4-fermion effective operator is valid. 

The next step is to calculate the relic density of neutrinos. Let us assume we are in a Universe where entropy is conserved so that we can define the co-moving number density $Y=n/s$ as before. If there is no entropy injected between freeze-out and today, then $Y_{\rm fo}=Y_{\rm today}=Y(T_\nu)$ and we only need to worry about the expansion of the Universe. For a relativistic species, $s=\frac{2\pi^2}{45}g_{*,S}(T)T^3$ and for a relativistic fermion $n=\frac{3\zeta(3)}{4\pi^2}g T^3$, where $g_{*,s}=(2+4\times 7/8)=11/2$ for photons and electrons, $g=2$ for a single fermionic species, and $\zeta(s)=\sum\limits_{n=1}^\infty\frac{1}{n^s}$ is the Riemann zeta function.

We can calculate the density of neutrinos from $\rho_{\nu,0}=m_\nu Y_{\rm fo}s_0$ and $\Omega_\nu\equiv \rho_\nu/\rho_c$. Using the expressions from above, we end up with
\begin{equation}
    \Omega_\nu h^2\simeq 0.12\times \frac{g}{g_{*,s}}\left(\frac{\sum m_\nu}{4.1{~\rm eV}}\right)\, .
\end{equation}
The current bound for neutrino mass is $m_\nu\lesssim 0.1$ eV, which leads to $\Omega_\nu h^2\lesssim 0.2\%$ for a single massive neutrino species today. 

\subsubsection{Relativistic relics}\label{sec:relativistic}
Now, let us consider a new particle $\chi$ that was relativistic at the time of decoupling. Let us also suppose that $T_{\rm fo}>T_\nu$ with instantaneous neutrino-decoupling at $T_\nu=1$ MeV. The number density of $\chi$ at $T_\nu$ is given by
\begin{equation}
    n_\chi(T=1{~\rm MeV})=\frac{135\zeta(3) g}{8\pi^4 g_{*,s}(T_{\rm fo})}s(T={\rm MeV}^+)\, ,
\end{equation}
where the ${\rm MeV}^+$ indicates the time leading up to neutrino-freezeout, and 
\begin{equation}
    Y_\chi=\frac{n_\chi(T=1{~\rm MeV})}{s_\gamma(T=1{~\rm MeV}^-)}=\frac{135\zeta(3) g}{8\pi^4 g_{*,s}(T_{\rm fo})}\frac{43}{22}\, .
\end{equation}
Here, ${\rm MeV}^-$ indicates the time right after neutrino-freezeout. This gives us
\begin{equation}
    \Omega_\chi h^2\simeq 0.12\times \frac{g}{g_{*,s}(T_{\rm fo})}\left(\frac{m_\chi}{\rm eV}\right)\, ,
\end{equation}
so freeze-out of a relativistic particle can populate the DM abundance if $m_\chi\sim 1-10$ eV with $g\sim 2, g_{*,s}(T_{\rm fo})\sim{\cal O}(10)$. However, earlier we saw that thermal DM should have $m_\chi\gtrsim$ keV, so to make these two statements consistent, we need for $g_{*,s}(T_{\rm fo})\gtrsim 1000$ for $m_\chi\gtrsim1$ keV, which is much more than the value in the SM! {Therefore, we can conclude that the freeze-out of relativistic species only works in non-standard cosmologies.} 

\subsubsection{Non-relativistic relics}\label{sec:nonrelativistic}
Does standard cosmology allow for non-relativistic relics? Let's find out. For a non-relativistic species, $n_\chi^{\rm eq}=g\left(\frac{m_\chi T}{2\pi}\right)^{3/2} e^{-m_\chi/T}$. As we will see, the exponential suppression will be key. 

Again, we apply the condition for freeze-out: $\Gamma = n_\chi^{\rm eq}\langle\sigma v\rangle=H$ to find
\begin{equation}
    n_\chi^{\rm eq}|_{\rm fo}\simeq g\left(\frac{m_\chi T_{\rm fo}}{2\pi}\right)^{3/2}e^{-m_\chi/T_{\rm fo}}=1.66\sqrt{g_*}\frac{T_{\rm fo}^2}{M_{\rm Pl}}\cdot\frac{1}{\langle\sigma v\rangle}\, ,
\end{equation}
and
\begin{equation}
    Y_{\rm fo}=\frac{n_\chi^{\rm eq}}{s}=\frac{H}{\langle\sigma v\rangle s}\simeq\frac{\sqrt{g_*}}{g_{*,s}}\frac{1}{\langle\sigma v\rangle T_{\rm fo}M_{\rm Pl}}\, .
\end{equation}
We see that $Y_{\rm fo}\ll 1$ is possible if $\langle\sigma v\rangle$ is sufficiently large. If $\langle\sigma v\rangle$ is large, then the DM interactions last longer and we end up with a smaller $n_\chi^{\rm eq}$. Since $n_\chi^{\rm eq}$ drops exponentially when the temperature $T<m_\chi$, we can estimate that $T_{\rm fo}\lesssim m_\chi$, but not too much below. 

To calculate the final DM abundance, we use
\begin{equation}
      \Omega_\chi=\frac{m_\chi Y_\chi s}{\rho_c}\sim \sqrt{g_*}\frac{x_f}{\langle\sigma v\rangle}\frac{1}{M_{\rm Pl}}\frac{s}{\rho_c}\, , 
\end{equation}
where we have defined $x_f\equiv m_\chi/T_{\rm fo}$.
Plugging in $s_{\rm today}=2891$ cm$^{-3},~\rho_c=1.05\times 10^{-5}h^2$GeV cm$^{-3}$, and $h=0.68$, we find for $\Omega_\chi h^2\simeq 0.12$ we need
\begin{equation}
    \langle\sigma v\rangle \sim \sqrt{g_*}x_f\frac{2.44\times 10^{-10}{\rm GeV}^{-2}}{\Omega_\chi}\simeq 10^{-9}{\rm GeV}^{-2}\, ,
\end{equation}
where we have used $x_f\sim 10$. 
Note that this is the {\it minimum} annihilation cross-section needed for a thermal candidate in order to avoid overabundance. We can also translate this cross-section into a benchmark for indirect searches: $\langle \sigma v\rangle\simeq 10^{-26}$ cm$^3/s$. 

Now that we have a rough-estimate for the required cross-section, we can ask what types of models give us such a cross-section? 
\begin{figure}
\centering
\includegraphics[width=0.4\textwidth]{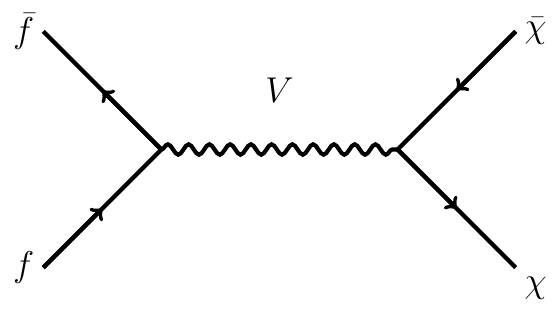}
\includegraphics[width=0.35\textwidth]{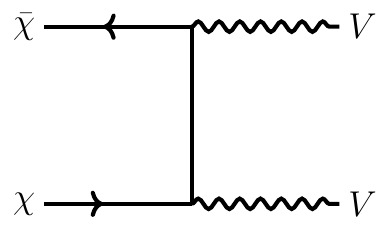}
\caption{({\bf left}) Example of an interaction between two SM fermions, $f$ and $\bar f$, with two fermionic DM particles $\chi$ and $\bar\chi$, which proceeds through a vector boson $V$. ({\bf right}) Example of a second possible annihilation channel for SM fermions, $f$ and $\bar f$ into two vector bosons $V$. }
\label{fig:annihilation}
\end{figure}
The annihilation cross-section for a particle of mass $m_\chi$ and weak-scale interactions will have $\sigma v\sim\frac{\alpha^2\pi}{m_\chi^2}$. Comparing this to the desired number above, we find that $m_\chi\sim 300$ GeV, which corresponds to the weak-scale. Therefore, a weak-scale DM candidate with weak-scale interactions can give us the observed relic abundance -- this coincidence is known as the ``WIMP miracle." 

Let us make a more generic statement. Suppose the annihilation goes through an $s$-channel process as shown in the left panel of Fig.~\ref{fig:annihilation}. 
Then, 
\begin{equation}
    \langle\sigma v\rangle\simeq\frac{|{\cal M}|^2}{32\pi m_\chi^2}
    \simeq \frac{g_\chi^2 g_f^2 m_\chi^2}{\pi(s-m_V^2)^2}
    =\frac{16\pi\alpha_\chi \alpha_f m_\chi^2}{(s-m_V^2)^2}\, .
\end{equation}
Let's consider the two cases:
\begin{enumerate}
    \item $m_\chi<m_V:\langle\sigma v\rangle\simeq \frac{16\pi\alpha_\chi \alpha_f m_\chi^2}{m_V^4}$.
    \item $m_\chi>m_V:\langle\sigma v\rangle_{\chi\bar\chi\to f\bar f}\simeq \frac{\pi\alpha_\chi \alpha_f}{m_\chi^2}$. In this scenario, we also need to include the process $\chi\bar\chi\to VV$, as seen in the right panel of Fig.~\ref{fig:annihilation}. 
    \begin{equation}
        \langle\sigma v\rangle_{\chi\bar\chi\to VV}\simeq \frac{\pi\alpha_\chi^2}{m_\chi^2}.
    \end{equation}
    If $\alpha_\chi\gg \alpha_f$, then this process dominates and we have a ``secluded" dark sector. Note that this scenario has no dependence on $\alpha_f$ and therefore is difficult to probe in a lab. However, there are indirect probes from CMB.  
\end{enumerate}
In either case, generally we can bound 
\begin{equation}
    \langle\sigma v\rangle\lesssim\frac{\pi{\rm max}(\alpha_\chi\alpha_f,\alpha_\chi^2)
    }{m_\chi^2}\, .
\end{equation}
We can take some lessons from this expression:
\begin{itemize}
    \item Let's consider weak-scale couplings ($\alpha_{\chi,f}\sim 10^{-2}-10^{-1}$). Substituting these into the annihilation cross-section formula and equating it to the required value gives $\langle\sigma v\rangle\simeq \alpha_W^2/(500{~\rm GeV})^2$. Here, both the couplings and mediator mass naturally fall within the weak scale. This {\it WIMP miracle} explains why weakly interacting massive particles (WIMPs) can naturally produce the observed DM abundance..
    \item Requiring perturbative couplings, ($\alpha_{\chi,f}\sim 1$) imposes an upper limit on the DM mass $m_\chi\lesssim 50-100$ TeV. This constraint is known as the {\it perturbative unitarity bound}~\cite{Griest:1989wd}. 
    \item Suppose $m_V\simeq$ 100 GeV. Then
    \begin{equation}
        \langle\sigma v\rangle\simeq\frac{m_\chi^2}{{\rm GeV}^2}\frac{\alpha_\chi\alpha_f}{\alpha_W^2}\frac{1}{10^9{~\rm GeV}^2}\, .
    \end{equation}
    For $\alpha_{\chi,f}\sim\alpha_W, \langle\sigma v\rangle$ drops below the necessary value when $m_\chi<$ GeV in scale, resulting in an overabundance of DM! Therefore, {\it new} mediators below the weak-scale are needed for sub-GeV DM candidates!\footnote{Millicharged DM is an exception.} This is known as the {\it Lee-Weinberg bound} and can also be used to constrain the mass of neutrinos~\cite{Lee:1977ua,Kolb:1986nf}
\end{itemize}

The WIMP miracle is indeed quite miraculous and has motivated much of the DM research over the last few decades. However, there were several assumptions baked into the discussion of the WIMP miracle. These include:
\begin{itemize}
    \item there is no chemical potential
    \item there are no resonances or threshold behavior
    \item annihilations of DM DM$\to$ SM SM interactions dominate
    \item annihilations happen during the radiation-dominated era of standard cosmology
    \item our Universe is homogeneous and isotropic
    \item co-moving entropy is conserved
\end{itemize} 

Modifying any of these assumptions can lead to significantly different conclusions. While the WIMP miracle provides strong motivation for model-building and experimental searches, it should not be viewed as a strict requirement.  
\subsection{No: Freeze-in}\label{sec:freezein}
If the answer to the question ``Was the DM ever in thermal equilibrium with the Standard Model" is no, then how was it populated? One explanation is that the DM has feeble couplings to us. Examples of such candidates are the sterile neutrino~\cite{Dasgupta:2021ies}, SuperWIMPs~\cite{Feng:2003xh,Feng:2003uy}, and freeze-in production~\cite{Hall:2009bx,Bernal:2017kxu}. Let's focus on this last point. 
Freeze-in DM is a class of DM in which the dark sector gets populated through the decay or annihilation of SM particles until the number density of the corresponding particle species becomes Boltzmann-suppressed. 

As a specific example, let's consider $s$-channel annihilations of $e^+e^-\to \chi\chi$, which proceeds through the left hand diagram in Fig.~\ref{fig:annihilation}. We can break this example down into two scenarios which depend on the relative mass of the mediator, $m_V$, to the DM mass $m_\chi$. 

{\bf IR-dominated:} The first case is when $m_V\ll m_\chi$. This scenario is known as ``IR-dominated" freeze-in. For the same $s$-channel annihilation, we have
\begin{equation}
    \langle\sigma v\rangle_{e^+e^-\to\chi\chi}\simeq\frac{\alpha_\chi\alpha_e}{T^2}\, ,
\end{equation}
and
\begin{equation}
    Y_\chi=\frac{n_\chi}{s}\simeq\frac{\Gamma}{g_{*,s}H}\simeq\frac{\alpha_\chi\alpha_e M_{\rm {Pl}}}{\sqrt{g_*}g_{*,s}T}\, .
\end{equation}
From this expression, we can take some lessons about IR freeze-in:
\begin{itemize}
    \item most of the abundance is produced at the lowest $T$ where the process is kinematically accessible, \ie $T=\max(m_\chi,m_e)$. When $T<m_\chi$, the DM becomes too heavy while when $T<m_e$, the electrons become too dilute. 
    \item One can estimate the couplings required by setting $Y_\chi(T_{\rm low})=Y_{\rm fo}$ (Eq.~\ref{eq:Ytoday}). There are two possibilities which depend on the relative size of the DM mass to electron mass:
    \begin{equation}
        \alpha_\chi\alpha_e\simeq\begin{cases}
        \sqrt{g_*}g_{*,s}|_{T=m_\chi}\times\frac{\rm eV}{M_{\rm Pl}}\simeq 3\times 10^{-27}& m_\chi > m_e\\
        \sqrt{g_*}g_{*,s}|_{T=m_e}\times\frac{\rm eV}{M_{\rm Pl}}\simeq 3\times 10^{-27}\frac{m_e}{m_\chi}& m_\chi < m_e
        \end{cases}
    \end{equation}
\end{itemize}
The couplings in this scenario are extremely small, far below the weak scale (recall $\alpha_W\sim 10^{-2}$). While testing models with such tiny couplings is challenging, detection may still be possible through direct or indirect searches, especially when the mediator mass is significantly smaller than the DM mass.

Another example of IR-dominated freeze-in is in the decay of a SM ``bath" particle, $B_1\to B_2\chi$, which is governed by the operator $\lambda B_1 B_2\chi$. Here, $B_2$ is a second SM particle and $m_{B_1}>m_{B_2}+m_\chi$. 
At high temperatures $T\gg m_{B_1}$, we have
\beq
Y_{B_1\to B_2\chi}(T)\sim \frac{M_{\rm Pl} m_{B_1}\Gamma_{B_1}}{T^3}\, .
\eeq
Note that due to the $T^{-3}$ behavior, this is strongly IR dominated and the relevant temperature is $T=m_{B_1}$. One can be more precise and solve the Boltzmann equations to get
\beq
Y_{B_1\to B_2\chi}\simeq \frac{135 g_{B_1}}{8\pi^3(1.66)g_{*,s}\sqrt{g_*}}\left(\frac{M_{\rm Pl}\Gamma_{B_1}}{m_{B_1}^2}\right)
\eeq

{\bf UV-dominated:} The second case is where $m_V\gg m_\chi$. This is known as ``UV-dominated" freeze-in and the production cross-section depends on a high scale $\Lambda$. If the DM is a scalar, we can write down a dimension-5 effective operator that governs the annihilation:\footnote{This operator as written is not $SU(2)_L$ invariant, but secretly $\lambda=v_H/\Lambda$.}
\begin{equation}
    \frac{g_\chi g_e \lambda}{\Lambda}\chi^2 e\bar e\to \langle\sigma v\rangle\simeq \frac{\alpha_\chi \alpha_e \lambda^2}{\Lambda^2}\, ,
\end{equation}
and the rate of producing a DM particle per electron is 
\begin{equation}
    \Gamma_{e^+e^-\to \chi\chi}=n_e\langle\sigma v\rangle\sim\frac{\alpha_\chi\alpha_e T^3\lambda^2}{\Lambda^2}~\textrm{for}~T\gg m_e, m_\chi\, . 
\end{equation}
The final DM abundance is then given by 
\begin{equation}
    Y_\chi=\frac{n_\chi}{s}=\frac{n_e\Gamma H^{-1}}{s}\simeq \frac{\Gamma}{g_{*,s}H}\simeq\frac{\alpha_\chi\alpha_e \lambda^2 M_{\rm Pl}T_{\rm RH}}{\sqrt{g_*}g_{*,s}\Lambda^2}\, .
\end{equation}
This expression gives us a few lessons about UV freeze-in:
\begin{itemize}
    \item The greatest abundance is produced at high $T~(T<\Lambda)$, and so the relic density is sensitive to the reheating of DM and the maximum available temperature ($T_{\rm RH}$). 
    \item A consequence of the above point is that laboratory detection is hard. If we assume that $m_\chi>$ keV, $T>$ MeV, then $\Lambda/(\alpha_\chi\alpha_e)^{1/4}\gtrsim10^6$ GeV. Such high-scales (or low-couplings) are difficult to detect. 
\end{itemize}
In general, UV freeze-in occurs when the operator which connects the DM to the SM is non-renormalizable,\footnote{If the freeze-in production arises from a renormalizable interaction, then the yield only depends on the coupling and particle masses, as we saw in the IR freeze-in section~\cite{Hall:2009bx}. } and the final relic abundance depends on the dimension of the operator. Consider SM SM$\to $ DM DM production through a generic, non-renormalizable operator with dimension $d=4+n$,
\begin{equation}
{\cal L}\supset \frac{1}{\Lambda^n}{\cal O}_{\rm SM}{\cal O}_{\rm DM}\, ,
\end{equation}
where ${\cal O}_{\rm SM,DM}$ are the SM and DM operators, respectively. From dimensional analysis, the production rate of the DM scales as $\Gamma\propto T^{2n+4}/\Lambda^{2n}$ so the relic density 
\beq
\Omega_\chi h^2\propto m_{\chi}Y_\chi\sim m_\chi \left(\frac{T_{\rm RH}^{2n-1}}{\Lambda^{2n}}\right)\, ,
\eeq
 where $T_{\rm RH}$ is the DM reheat temperature. Thus, for a dimension-5 operator ($n=1$), we get $\Omega_\chi\sim T_{\rm RH}/\Lambda^2$, as we saw before, while a dimension-6 operator ($n=2$) would give $\Omega_\chi\sim T_{\rm RH}^3/\Lambda^4$. More detailed discussion can be found in Ref.~\cite{Elahi:2014fsa}.

\subsection{No: Misalignment mechanism}\label{sec:misalignment}

The misalignment mechanism is a compelling theory for producing DM in the early Universe, particularly for axions and axion-like particles. This mechanism has also been proposed for vector particles.

In the early Universe, the scalar field associated with axions or similar particles could have started in a random position due to quantum fluctuations. As the Universe expanded and cooled, the dynamics of the field would lead it to evolve towards the minimum of its potential. However, if the field initially starts misaligned with the minimum, it will oscillate around this minimum as the Universe continues to expand. 

The energy of these oscillations can be converted into particles, effectively producing DM. The critical aspect of this mechanism is that the abundance of DM produced is dependent on the initial value of the field, which can vary widely due to the random quantum fluctuations. As a result, the misalignment mechanism provides a natural and compelling way to account for the observed density of DM in the Universe, contingent on the properties of the scalar field, such as its mass and coupling constants \cite{Preskill:1982cy, Abbott:1982af, Dine:1982ah}.

Moreover, the misalignment mechanism predicts a characteristic density profile for the DM produced, leading to unique signatures that can be searched for in various astrophysical and cosmological observations. The misalignment mechanism, and more generally discussion of axions and other ultralight DM candidates, deserve a lecture of their own and we will not discuss them further in these notes. The interested reader can find more discussion in references such as Refs.~\cite{Marsh:2015xka,Chadha-Day:2021szb,Antypas:2022asj}

\subsection{Caveats}\label{sec:caveats}
In the previous sections, we have utilized the cosmological production of DM as a guiding framework for identifying potential avenues for DM detection. The term ``guiding principle" is crucial here, as there are important caveats to relying solely on cosmological production as a metric for exploration.

One significant consideration is the uncertainty surrounding the early Universe. Prior to Big Bang Nucleosynthesis (BBN), much of our understanding of the physics during this epoch is based on extrapolations from the Standard Model and elements like DM. Phenomena such as inflation, the existence of DM, and baryon asymmetry all suggest that crucial components of our theoretical framework may be absent during these early times, rendering our extrapolations uncertain. For example,
\begin{itemize}
\item there could exist some other dark particle that could decay into DM, increasing the DM abundance. 
\item some other particle could decay into SM particles, diluting the DM abundance.
\item our Universe could have undergone a period of early-matter domination, therefore changing our cosmological history.
\end{itemize}
However, this uncertainty can also be advantageous. For instance, gaining insights into the annihilation cross-section could validate the WIMP miracle and deepen our comprehension of the Universe's history.

We end this section with the key lessons about producing DM:
\begin{itemize}
    \item Although our knowledge of the fundamental nature of DM is limited, formulating a coherent theory is significantly constrained.
    \item Once a theoretical framework for DM is established, we can investigate how it may have been produced in the early Universe.
    \item The Boltzmann equation provides a means to track the evolution of number density, potentially leading to the correct abundance in various scenarios.
    \item Nonetheless, we must exercise caution in utilizing relic abundance to inform the parameter space of our models; if our assumptions about the Universe's history are flawed, we risk pursuing misleading directions.
\end{itemize}
\section{Looking for Dark Matter}\label{sec:searches}

Now that we have established the range of possible DM candidates and outlined key guiding principles for prioritizing the parameter space of models, we can now turn to the crucial question: how do we search for DM? The method we employ depends on the mass of the DM candidate and the nature of its interactions with the Standard Model. Given the vast array of proposed DM candidates -- ranging from ultralight axions to WIMPs and even macroscopic objects -- it is clear that no single search strategy will suffice. Rather, a diverse and complementary set of approaches is essential. This principle was strongly emphasized during the Snowmass 2021 process~\cite{Chou:2022luk}, encapsulated in the call to ``Delve Deep, Search Wide, Aim High." 

Broadly speaking, DM searches can be categorized into four main approaches:  

\begin{enumerate}
\item {\bf Collider and accelerator-based searches} --  These aim to produce DM or more generally dark sector particles in high-energy collisions, such as at the Large Hadron Collider (LHC) or fixed-target experiments, and search for missing energy signatures or the decay products of dark sector mediators.  
\item {\bf Indirect detection} -- This method looks for signals of DM annihilation or decay in astrophysical environments, searching for excess gamma rays, cosmic rays, or neutrinos.  
\item {\bf Direct detection} -- This approach seeks to measure DM interactions with ordinary matter in highly sensitive underground detectors, looking for nuclear or electron recoils.  
\item {\bf Astrophysical probes} -- These use cosmological and astrophysical observations to infer the presence and properties of DM through its gravitational effects on structure formation, the cosmic microwave background (CMB), and other large-scale phenomena. There are also smaller-scale probes that look for the imprints of DM interactions in stellar environments such as supernovae and neutron stars.
\end{enumerate}

The first three categories are sometimes colloquially summarized as ``make it, break it, or shake it." Colliders {\it make} DM, indirect detection experiments observe the {\it break}ing of DM by looking for energetic products of DM annihilation or decay, and direct detection looks for the {\it shake} of DM's interactions with normal matter. 

We are currently in an exciting era for DM searches, with new experimental and observational advances on the horizon. In these lectures, we began by discussing astrophysical probes, particularly focusing on constraints from the matter power spectrum. Looking ahead, there are numerous upcoming opportunities in this domain, including next-generation CMB experiments~\cite{CMBS4} and large-scale spectroscopic surveys~\cite{DESI,Spec-S5:2025uom}, which will provide new insights into DM properties.  

For light and ultralight DM candidates, innovative small-scale and pathfinder experiments are being actively developed, offering promising new ways to test theories in previously unexplored regions of parameter space. On the collider front, the High-Luminosity Large Hadron Collider (HL-LHC) will push the energy and precision frontier, while R\&D for future colliders holds exciting potential for extending our discovery reach.  

Each of these approaches provides a unique and necessary perspective on the DM puzzle, and together, they form a comprehensive strategy. The interplay between different search methods ensures that we are well-positioned to explore the full range of possibilities. In the coming years, continued progress across these diverse search strategies will be key to unraveling the mystery of dark matter. Underpinning all of these efforts is theory, which plays a central role in guiding experimental priorities, informing design, interpreting results, and charting paths into uncharted territory.

As a specific example, we will focus in this section on the third probe, direct detection. We will begin with a brief overview of DM-nuclear scattering before moving on to techniques to search for sub-GeV DM candidates. 
\subsection{Direct Detection}\label{sec:DD}

Let us now shift our focus to the direct detection of DM, specifically in the context of thermally produced DM. Direct detection experiments play a crucial role in the search for DM, offering a unique and complementary approach to collider and indirect detection strategies. Several key features make direct detection a particularly powerful probe:  

\begin{itemize}
    \item \textbf{Direct access to the galactic DM halo}: Unlike collider searches, which can produce new states that may or may not correspond to the actual cosmological DM population, direct detection experiments probe the local DM halo that permeates our galaxy. This direct interaction with ambient DM allows for a more straightforward test of its properties.  

    \item \textbf{Adaptability and responsiveness}: Direct detection experiments can quickly respond to new signals or hints of DM. If a potential signal is observed in one experiment, similar detectors can be reconfigured to perform follow-up measurements, helping to confirm or refute the discovery.  

    \item \textbf{Model independence}: Unlike collider searches, which often rely on specific production mechanisms or decay chains, direct detection experiments can simultaneously search for multiple potential DM signatures across a wide range of DM masses. This broad sensitivity makes them a versatile tool for exploring different theoretical possibilities.  

    \item \textbf{Controlled and well-characterized environments}: These experiments operate in carefully designed, low-background conditions, allowing for precise control over systematic uncertainties. This clean and configurable setting enables detailed studies of potential backgrounds, ensuring that any observed excess events are thoroughly vetted for DM-like characteristics.  
\end{itemize}

The fundamental observable in direct detection experiments is the event rate, which represents the number of scattering events per unit time per unit detector mass. The differential event rate is given by:  

\begin{equation}
    \frac{dR}{dE_R} \propto \sigma_i n_\chi n_T e^{-E_R/E_0}\, ,
\end{equation}

where:  

\begin{itemize}
    \item \( n_\chi \) and \( n_T \) are the number densities of DM and the target material, respectively,  
    \item \( \sigma_i \) represents the DM interaction cross-section with a Standard Model particle \( i \), which could be a nucleon, nucleus, or electron,  
    \item \( E_R \) is the recoil energy of the target, and  
    \item \( E_0 \) is a characteristic energy scale set by the velocity distribution of DM particles.  
\end{itemize}

From this expression, we can infer key factors that influence the event rate:  

\begin{itemize}
    \item {Larger cross-sections} (\( \sigma_i \)) increase the probability of interaction, making it more likely that a DM particle will scatter within the detector.  
    \item {Higher number densities} (\( n_\chi, n_T \)) lead to more frequent interactions, enhancing the overall event rate.  
    \item {Lower recoil energies} (\( E_R \)) are generally favored due to the exponential suppression at high energies, meaning that many direct detection experiments are optimized to search for low-energy recoils.  
\end{itemize}

\begin{figure}
    \centering
    \includegraphics[width=0.75\linewidth]{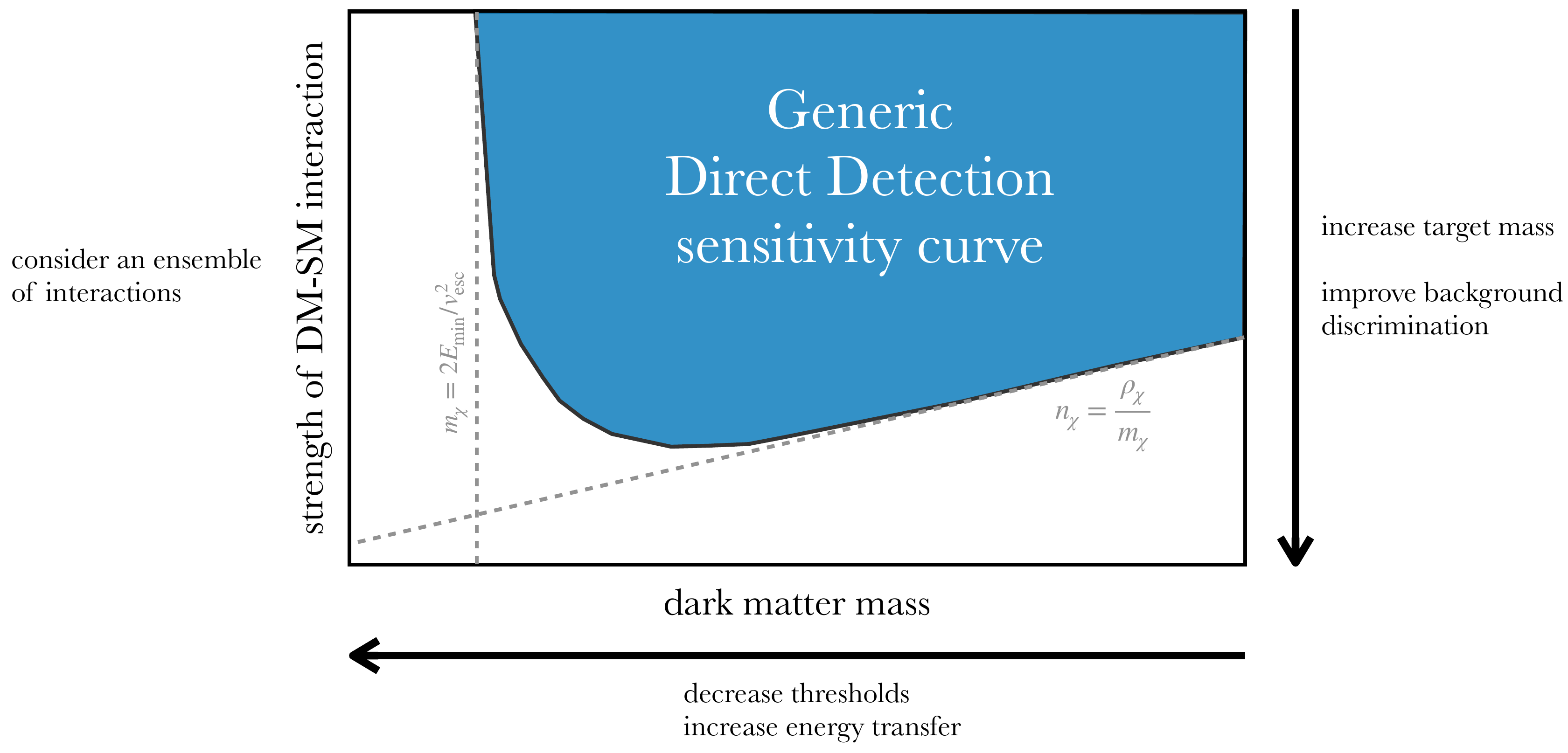}
    \caption{Schematic representation of an experimental direct detection result. The shaded, blue region shows the sensitivity or constraining power of an experiment on the DM-SM scattering cross-section as a function of the DM mass. A major goal for direct detection searches is to fully explore this space. One can push the blue region down, to lower cross-sections, by increasing the target mass and/or improving background discrimination. Decreasing the energy threshold or considering processes that increase the energy transfer can push the reach to lighter DM masses. Finally, one should consider a variety of different DM-SM interactions.}
    \label{fig:generic_DD_curve}
\end{figure}

In Fig.~\ref{fig:generic_DD_curve}, we present a schematic representation of the direct detection landscape, illustrating the sensitivity reach of experiments in the search for dark matter (DM). The shaded blue region denotes the sensitivity or constraining power of a given experiment on the DM-SM interaction, which depends on the DM mass \( m_\chi \).  

There are some universal trends that all direct detection experiments share:  

\begin{itemize}
    \item \textbf{High-mass behavior:} At large DM masses, the number density of DM particles decreases as \( n_\chi \propto m_\chi^{-1} \), following from the fact that the total DM energy density remains fixed. Since the event rate scales with the number density, the sensitivity of an experiment similarly decreases as \( m_\chi^{-1} \). This means that as the DM mass increases, fewer particles are available to interact with the detector, making detection more challenging.  

    \item \textbf{Low-mass behavior:} At small DM masses, the experiment eventually encounters a kinematic threshold determined by the minimum energy required to produce a detectable recoil signal, denoted as \( E_{\rm min} \). If the DM mass is too low, the available kinetic energy in the DM-nucleus (or DM-electron) interaction becomes insufficient to produce an observable event, leading to a sharp loss of sensitivity. This threshold is governed by both the detector technology and the type of target material used.  
\end{itemize}  

The overarching goal of direct detection experiments is to probe as much of the DM parameter space as possible by pushing the boundaries along different axes. There are several key strategies to enhance sensitivity:  

\begin{itemize}
    \item \textbf{Probing smaller interaction cross-sections (\(\sigma\)):} To access weaker interactions, experiments can either increase the detector's target mass, thereby increasing the number of potential scattering events, or improve background discrimination techniques. Reducing background noise allows experiments to identify smaller signals that might otherwise be lost in the data.  

    \item \textbf{Enhancing sensitivity to low-mass DM:} One approach to access lower DM mass is to lower the detector’s energy threshold, enabling the detection of even smaller energy transfers. Alternatively, one can explore processes that enhance energy deposition, such as collective excitations in materials, inelastic scattering, or utilizing different target materials with lighter nuclei, which result in larger recoil energies for a given DM velocity.  

    \item \textbf{Considering an ensemble of interactions:} Traditional searches focus on spin-independent and spin-dependent nuclear scattering, but exploring alternative interactions such as electron recoils, magnetic dipole interactions, or other dark sector-mediated interactions can open new avenues for detection. Different interaction channels can provide complementary constraints and increase overall discovery potential.  
\end{itemize}

\subsubsection{DM-nuclear scattering}\label{sec:DMnscattering}
Traditionally, the community has focused on DM-{\it nuclear} scattering, which is well-suited for WIMP candidates. To see why, let's take a look at the kinematics. 

\begin{figure}
    \centering
    \includegraphics[width=0.25\linewidth]{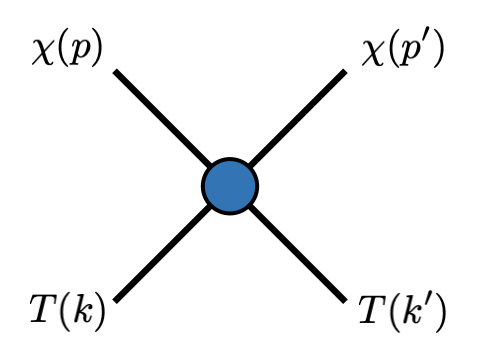}
    \caption{Definition of the kinematic variables for the direct detection of DM, where $\chi$ is the DM candidate and $T$ is the target. Primed ($'$) variables denote final states.}
    \label{fig:kinematics}
\end{figure}

In Figure~\ref{fig:kinematics}, we define the momentum for the incoming and outgoing DM as $p,p'$ respectively, and the momentum for the incoming and outgoing target as $\vec k, \vec k'$ respectively. Therefore, the momentum transfer is given by $\vec q\equiv \vec p-\vec p'=\vec k'-\vec k$. For DM-nuclear scattering, we can treat the nucleus as a particle at rest, so $\vec k=0$. 
The initial and final energies of the system are given by 
\begin{align}
    E_i&=\frac{|\vec p|^2}{2m_\chi}=\frac{1}{2}m_\chi v_\chi^2\nonumber\\
    E_f&=\frac{|\vec p-\vec q|^2}{2m_\chi}+\frac{|\vec q|^2}{2m_N}\, ,
\end{align}
where the second term in the second line comes from the recoil energy of the nucleus. Energy conservation gives us the relationship
\begin{equation}
    \frac{\vec p\cdot\vec q}{m_\chi}=\frac{|\vec q|^2}{2\mu_{\chi N}}\to q_{\rm max}=\frac{2\mu_{\chi N}|\vec p|}{m_\chi}=2\mu_{\chi N}v_\chi\, .
\end{equation}
From this, we can then calculate the maximum recoil energy of the nucleus,
\begin{equation}
\label{eq:DDn_energy}
    E_{R,{\rm max}}=\frac{q_{\rm max}^2}{2m_N}=2\frac{\mu_{\chi N}^2}{m_N}v_\chi^2\simeq~{\rm eV}\times\left(\frac{m_\chi}{100{~\rm MeV}}\right)^2\left(\frac{20{~\rm GeV}}{m_N}\right)\,.
\end{equation}
So, if we have a typical WIMP with $m_\chi\sim 100$ GeV and a nuclear target of $m_N\sim 131$ GeV, then we get recoil energies of $E_R\sim 150$ keV. This number gives us a good benchmark for experimental thresholds. 

\begin{figure}
    \centering
    \includegraphics[width=0.45\linewidth]{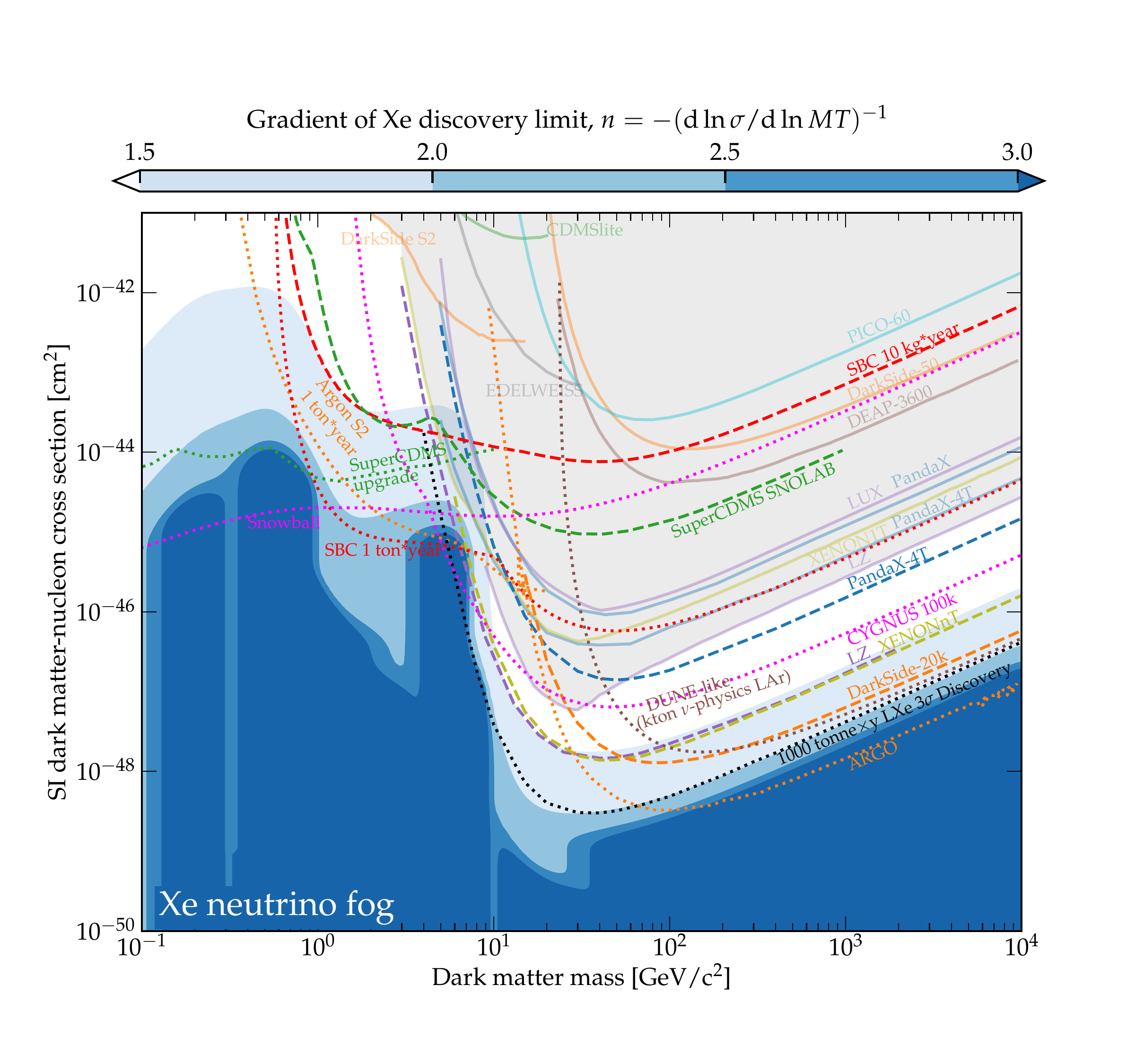}\hspace{-4mm}
    \includegraphics[width=0.55\linewidth]{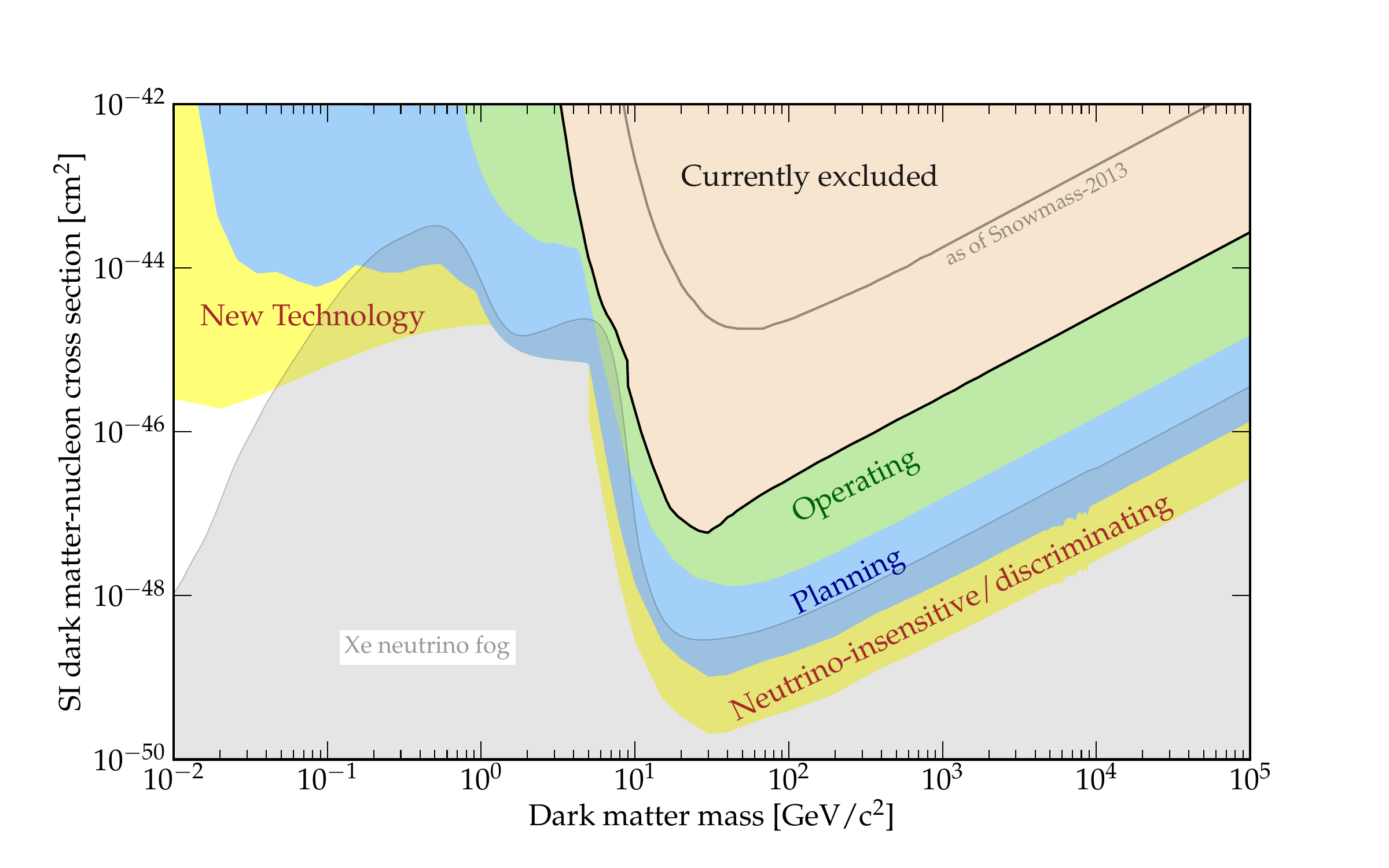}
    \caption{({\bf left}) Status of the direct detection spin-independent DM-nuclear scattering as of 2021. 
        ({\bf right}) future sensitivity. Figures used with permission from the 2021 Snowmass Proceedings~\cite{Akerib:2022ort,Cooley:2022ufh}.}
    \label{fig:DDN_SI}
\end{figure}

\begin{figure}
    \centering
    \includegraphics[width=0.45\linewidth]{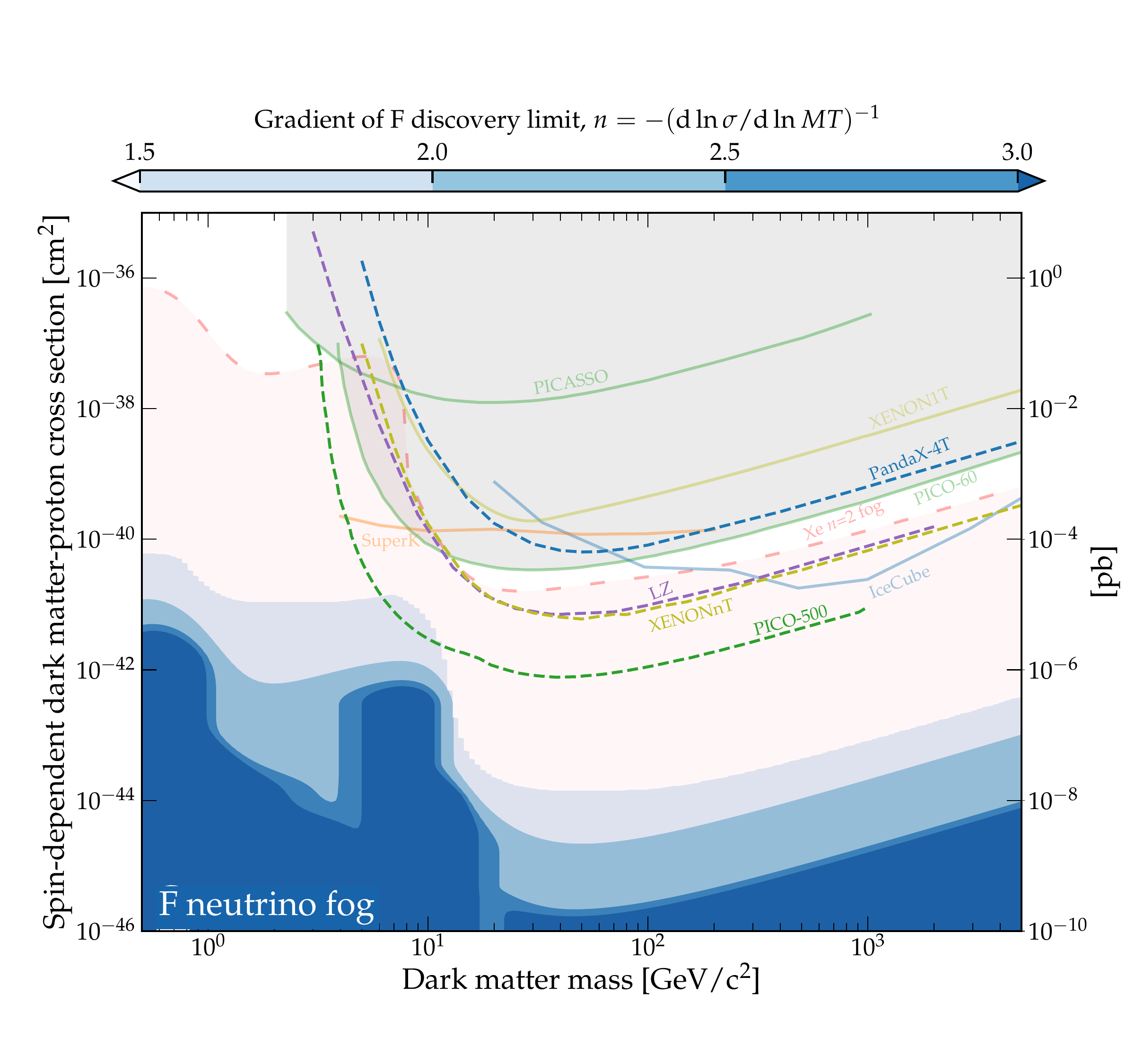}
    \includegraphics[width=0.45\linewidth]{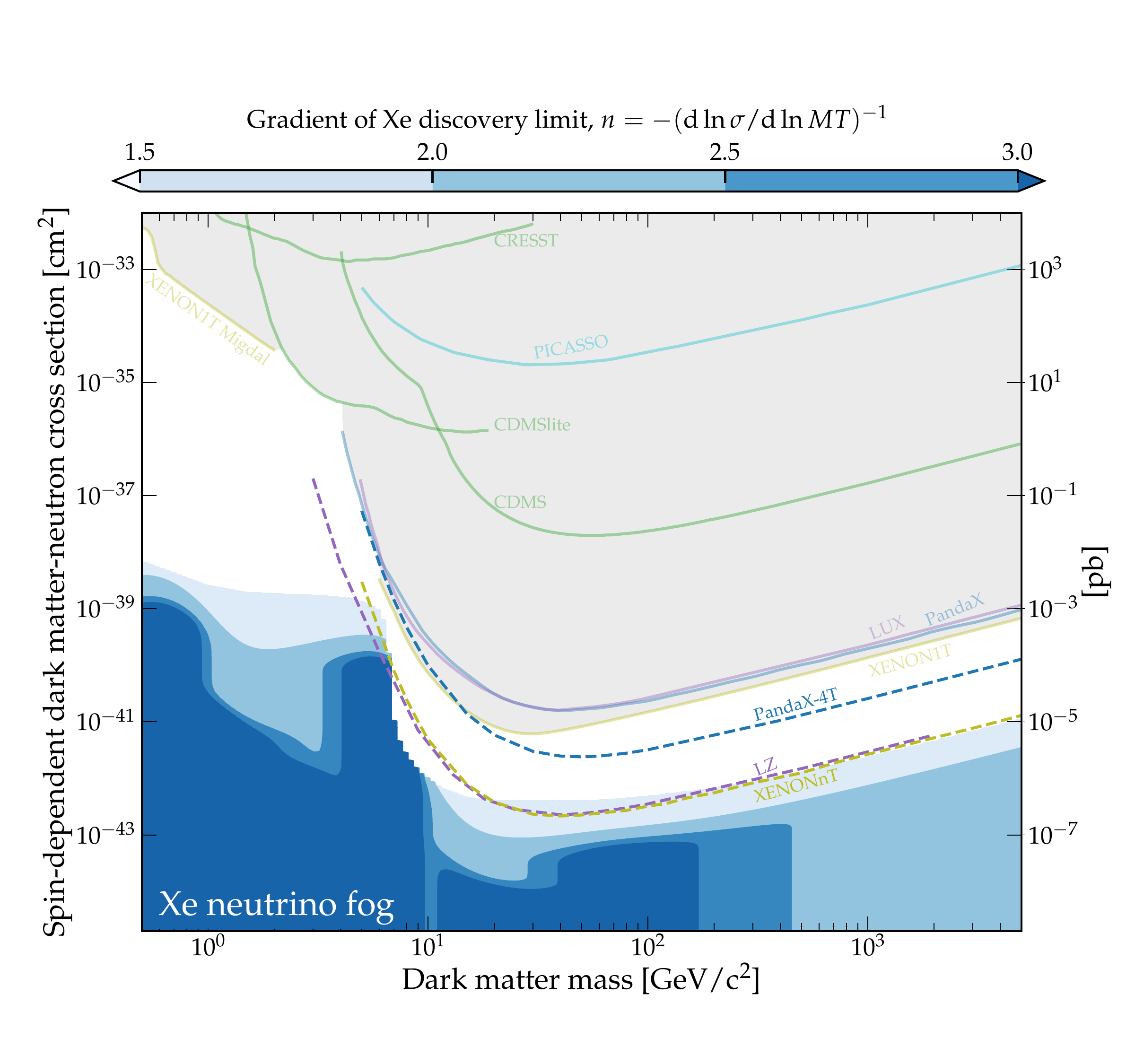}
    \includegraphics[width=0.65\linewidth]{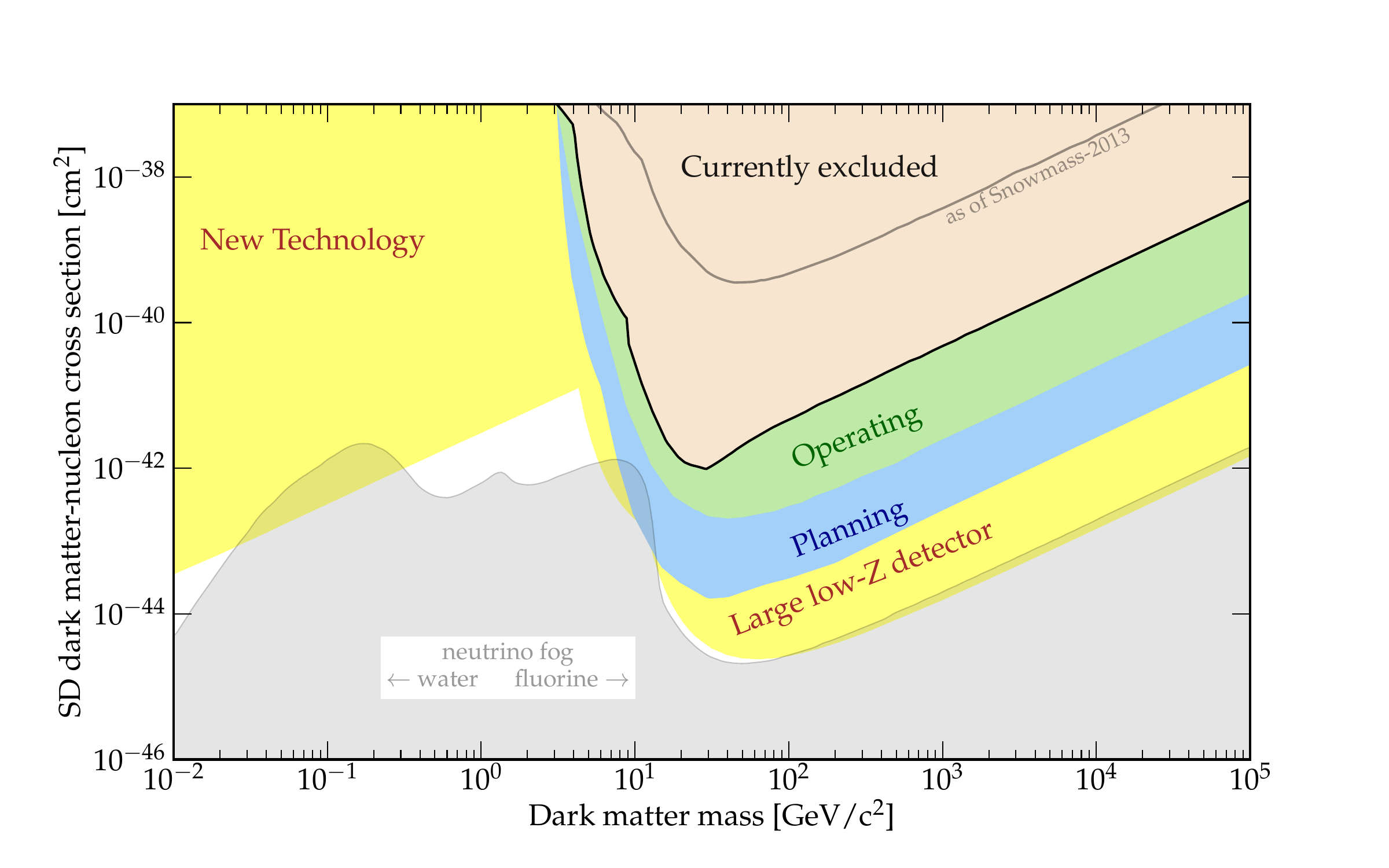}
    \caption{Status of the direct detection spin-dependent DM-nuclear scattering for ({\bf top, left}) proton and ({\bf top, right}) neutron couplings as of 2021. ({\bf bottom}) future sensitivity. Top figures courtesy of Ben Loer. Bottom figure used with permission from the 2021 Snowmass Proceedings~\cite{Cooley:2022ufh}. }
    \label{fig:DDN_SD}
\end{figure}

Now that we have some general numbers in mind, we can begin to calculate the rate for DM-nuclear scattering. The DM interactions are given at the microscopic level, \ie with quarks and gluons. Let's consider a fermionic DM candidate that is mediated by weak-scale mediators. We can write the interactions between the DM and the quarks through dimension-6 operators:
\begin{equation}
\frac{c_V}{m_Z^2}\bar\chi\gamma^\mu\chi\bar q\gamma_\mu q+\frac{c_A}{m_Z^2}\bar\chi\gamma^\mu\gamma^5\chi\bar q\gamma_\mu\gamma^5 q + \frac{c_S m_q}{m_h^2 v}\bar\chi\chi\bar qq+\ldots
\label{eq:dim6ops}
\end{equation}
Next, we need to match the quark-level operators to nucleon-level operators with a similar tensor structure, \ie $\bar q\gamma_\mu q\sim \bar u\gamma_\mu u$, etc. 

For example, the vector current which appears in the first term of Eq.~\ref{eq:dim6ops} becomes
\begin{equation}
    \langle n(k')|\bar q\gamma^\mu q|n(k)\rangle = \bar u_n(k')\left[F_1^{q,n}(q^2)\gamma^\mu+\frac{i}{2m_n}F_2^{q,n}(q^2)\sigma^{\mu\nu}q_\nu\right]u_n(k)\, ,
\end{equation}
where the $F_{1,2}$ are the matrix elements for the light quark operators in nucleon states. Note that this has the analogous form to the vector current in QED.
 In the $q^2\to 0$ limit, $F_1^{q,n}(0)$ represents the quark content of nucleon $n$. For example, $F_1^{u,p}(0)=2$ and $F_1^{d,p}(0)=1$. $F_2^{q,n}$ encodes the contribution of quark flavor $q$ to the nucleon anomalous magnetic moment and is typically subdominant.

Generally, we can write the DM-nucleus scattering cross-section as~\cite{Schumann:2019eaa}
\begin{equation}
\label{eq:DMnucleus}
    \frac{d\sigma}{dE_R}=\frac{m_N}{2 v^2\mu_{\chi N}^2}\left[\sigma_{\rm SI}F_{\rm SI}^2(q^2)+\sigma_{\rm SD}F_{\rm SD}^2(q^2)\right]\, ,
\end{equation}
where the rate is split into a spin-independent (SI) and spin-dependent (SD) contribution. At very small momentum transfer $q=\sqrt{2m_N E_R}$, the DM deBroglie wavelength is large and therefore the DM scatters coherently off of the entire nucleus. However, at higher momentum transfers, and therefore smaller DM deBroglie wavelengths, the interaction is no longer coherent. This loss of coherence is encoded in the nuclear form factors $F_{\rm SI, SD}$. 

For the spin-independent case, we can approximate the nuclear form-factor with the Helm form factor~\cite{Helm:1956},
\begin{equation}
    F_{\rm SI}(x)=\frac{3 j_1(x)}{x}\exp\left(-\frac{(xs)^2}{2R_N^2}\right)\, ,
\end{equation}
where $x\equiv q R_N, j_1(x)=\frac{\sin x}{x^2}-\frac{\cos x}{x}$, the effective nuclear radius $R_N\simeq 1.14 A^{1/3}$ fm, and the nuclear skin thickness $s\simeq 0.9$ fm~\cite{Lewin:1996QED}. Note that these last two quantities are nuclei dependent. We can also define
\begin{equation}
    \sigma_{\rm SI}=\sigma_n\frac{\mu_{\chi N}^2}{\mu_{\chi n}^2}\frac{\left[f_p Z+f_n(A-Z)\right]^2}{f_n^2}\, ,
\end{equation}
where $\sigma_n$ is now the DM-{\it nucleon} scattering cross-section, $f_{p,n}$ encodes any isospin-violation ($f_p\neq f_n$), and $A,Z$ are the atomic mass and number of the nucleus, respectively~\cite{Ellis:2008}. 

Note the $A^2$ dependence of the scattering rate; this is known as ``coherent" scattering and tells us that there are distinct advantages to using heavier elements (larger $A$) for spin-independent scattering. Having a variety of different targets is also desirable as one can then test the $A^2$ dependence of the scattering rate. 

In contrast to SI interactions, which benefit from the $A^2$ {coherent enhancement} due to all nucleons contributing in phase, {spin-dependent (SD) interactions} arise when DM couples to the {spin} of individual nucleons. Since nuclear spins are typically {dominated by unpaired nucleons}, SD interactions are often weaker and require different experimental strategies.

The differential cross-section for {DM-nucleus spin-dependent scattering} is given by 

\begin{equation}
\label{eq:diffSD}
    \frac{d\sigma_{\rm SD}}{d|\vec q|^2} = \frac{8}{\pi}\frac{G_F^2}{v^2} \frac{J (J+1)}{J} \left( a_p \langle S_p \rangle + a_n \langle S_n \rangle\right)^2\frac{S(|\vec q|)}{S(0)} \,,
\end{equation}
where \( G_F \) is the Fermi constant,  \( \mu_{\chi N} \) is the DM-nucleus reduced mass, \( J \) is the total nuclear spin, \( a_p \) and \( a_n \) are the {effective DM-proton and DM-neutron coupling constants}, and \( \langle S_{p,n} \rangle =\langle N|S_{p,n}|N\rangle\) are the {expectation values of the proton and neutron spin contributions} in the nucleus. $S(|\vec q|)$ is the spin-structure function, and can be obtained from detailed nuclear theory calculations using \eg {\it ab initio} techniques~\cite{PhysRevLett.128.072502}. 

Eq.~\ref{eq:diffSD} shows that SD interactions are {sensitive to the spin structure of the nucleus} rather than the total nucleon count.
Since {only nucleons with net spin contribute} to SD interactions, different nuclei have different sensitivities:  
\begin{itemize}
\item { $^{19}$F ($J=1/2$), $^{23}$Na ($J=3/2$), $^{27}$Al ($J=5/2$), $^{127}$I ($J=5/2$)} are sensitive to {proton-spin couplings} (\( a_p \)).  
\item {$^{129}$Xe ($J=1/2$), $^{131}$Xe ($J=3/2$), $^{29}$Si ($J=1/2$), $^{73}$Ge ($J=9/2$)} are sensitive to {neutron-spin couplings} (\( a_n \)).  
\end{itemize}
Notably, argon targets are insensitive to SD interactions while the spin-structure of fluorine, a relatively light nucleus ($Z=9$), makes it an excellent target for SD interactions.
Since SD interactions {do not benefit from coherent enhancement} (\( A^2 \) scaling), SD detection is generally more challenging than SI detection and the current cross-section constraints are roughly 5-6 orders of magnitude weaker. However, SD interactions provide a {complementary probe} of DM models, especially those with axial couplings.

Given the expressions for the differential cross-section, the DM-nucleus scattering rate can be calculated through
\begin{equation}
\frac{dR}{dE_R}=N_T\frac{\rho_\chi}{m_\chi}\int_{v_{\rm{min}}}^{v_{\rm{esc}}} v g_\chi(v)\frac{d\sigma}{dE_R} dv\, ,
\end{equation}
where $N_T$ is the number density of targets and $\rho_\chi\simeq 0.4$ GeV/cm$^3$ is the local DM density. $g_\chi(v)$ encodes the galactic DM velocity distribution. $v_{\rm min}=\sqrt{\frac{m_N E_R}{2\mu_{\chi N}^2}}$ is the minimum DM velocity needed to give an energy recoil $E_R$ while $v_{\rm esc}\simeq 544$ km/s is the escape velocity of the galaxy. $\mu_{\chi N}\equiv m_Nm_\chi/(m_N+m_\chi)$ is the reduced mass of the DM-nucleus system.
The status of spin-independent and spin-dependent searches, as of 2021, are shown in Figs.~\ref{fig:DDN_SI} and~\ref{fig:DDN_SD}, respectively. 

In future experiments, improved {low-threshold detectors} and {new detection techniques} (\eg quantum sensors, directional detectors) may further improve sensitivity to DM interactions, allowing for a {more complete exploration of the DM-nucleon parameter space}. 

\subsubsection{DM-electron scattering}\label{sec:DMescattering}
In the previous section, we focused on DM-nuclear interactions and saw in Eq.~\ref{eq:DDn_energy} that the recoil energies for a typical WIMP candidate of $m_\chi\sim 100$ GeV on a target of $m_N\sim 131$ GeV could give us ${\cal O}(100{~\rm keV})$ of energy. What about sub-GeV DM? If we use the same expression for a $m_\chi\sim 100$ MeV candidate, we find that $E_R\sim 0.1$ eV, which is not only significantly smaller than the WIMP case, but also well below the thresholds of any WIMP direct detection experiment. However, not all is lost. Instead of considering DM-nuclear interactions, let's look at DM-electron interactions, and specifically, at DM-electron scattering. 

The kinematics in this scenario are different as the assumption that $k=0$ no longer holds due to the motion of the electron, which is now the fastest particle in the system. The deposited energy from the DM scatter is given by
\beq
\Delta E_e\equiv E_e(\vec k')-E_e(\vec k)=\frac{\vec p\cdot\vec q}{m_\chi}-\frac{q^2}{2m_\chi}\, .
\eeq
Conservation of energy gives us
\begin{align}
\Delta E_E&=-\Delta E_\chi-\Delta E_N=-\frac{|m_\chi\vec v-\vec q|^2}{2m_\chi}+\frac{1}{2}m_\chi v^2-\frac{q^2}{2m_N}\nonumber\\
&=\vec q\cdot\vec v-\frac{q^2}{2\mu_{\chi N}}\, .
\end{align}
In practice, $\Delta E_N$ is very small so we can make the substitution $\mu_{\chi N}\to m_\chi$ and get
\beq
\label{eq:DeltaEe}
\Delta E_e=\vec q\cdot\vec v-\frac{q^2}{2m_\chi}\, .
\eeq
Note that in contrast to the nuclear scattering scenario, here an arbitrary-sized momentum transfer $q$ is possible. Therefore, we can calculate the maximum possible energy transfer by maximizing Eq.~\ref{eq:DeltaEe} with respect to $q$ to find
\beq
\Delta E_e^{\rm max}=\frac{1}{2}\mu_{\chi N}^2 v^2\simeq \frac{1}{2}{~\rm eV}\left(\frac{m_\chi}{{\rm MeV}}\right)\, .
\eeq
In short, all of the kinetic energy is available to excite the electron! Returning back to our DM candidate with $m_\chi=100$ MeV, we see that we now have $\Delta E_e\sim 50$ eV of energy to detect, as opposed to the 0.1 eV from the nuclear-scattering case. Although small, this is now at the level that experiments can detect. 

What are some typical values one can expect for $q$ and $E_e$? Since the electron is both the lightest and fastest particle in the system, its kinematics set the scales. The velocity of an electron in a bound state is given by $v_e\sim Z_{\rm eff}\alpha\sim 10^{-2}$ (contrast this to the DM velocity, which is $v_\chi\sim 10^{-3}$). Then, the typical momentum transfer is
\begin{align}
    q_{\rm typ}&\simeq \mu_{\chi e}v_{\rm rel}\simeq m_e v_e\sim Z_{\rm eff}\alpha m_e\\
    &\simeq Z_{\rm eff}\times 4{~\rm keV}\, .
\end{align}

Away from threshold, the first term of Eq.~\ref{eq:DeltaEe} dominates so $q\gtrsim \Delta E_e/v=\Delta E_e/(4 Z_{\rm eff}{\rm eV})\times q_{\rm typ}$ which leads to
\beq
\Delta E_e^{\rm typ}\sim {\rm eV}\, .
\eeq

Now we can discuss the actual calculation of the DM-electron scattering rate~\cite{Essig:2015cda}. 
If a DM particle scatters with an electron bound in a stationary state, such as within an atom or a crystal, it can excite the electron from an initial energy level \(1\) to a higher energy level \(2\). This occurs by transferring energy \(\Delta E_{1\to2}\) and momentum \(\vec{q}\) to the electron. 

We can treat the atom as an infinitely heavy object that absorbs momentum without recoiling, which is an excellent approximation (\(<1\%\) error) for momentum transfers of interest, typically on the order of keV.

The cross section for free \(2 \to 2\) scattering is expressed as:
\begin{equation}
\sigma v_{\rm free} = \frac{1}{4 E_\chi' E_e'} \int \frac{d^3 q}{(2\pi)^3} \frac{d^3 k'}{(2\pi)^3} \frac{1}{4 E_\chi E_e} (2 \pi)^4 \delta(E_i - E_f) \delta^3(\vec k + \vec q - \vec k') \overline{|\mathcal M_{\rm free}(\vec q\,)|^2} \, ,
\end{equation}
where \(\mathcal{M}_{\rm free}\) is the matrix element, and \(\overline{|\mathcal M|^2}\) represents its squared magnitude, averaged over initial and summed over final spins.

For an unbound electron, the non-relativistic scattering amplitude is:
\begin{equation}
\langle \chi_{\vec p - \vec q \,}, e_{\vec k'} | H_{\rm int} | \chi_{\vec p}, e_{\vec k} \rangle
= C \, \mathcal{M}_{\rm free}(\vec q \,) \times (2 \pi)^3 \delta^3(\vec k - \vec q - \vec k') \, ,
\end{equation}
where \(H_{\rm int}\) is the interaction Hamiltonian, and \(C\) is a constant. For a bound electron, the amplitude is modified to:
\begin{equation}
\langle \chi_{\vec p - \vec q \,}, e_2| H_{\rm int} | \chi_{\vec p}, e_1 \rangle =
C \, \mathcal{M}_{\rm free}(\vec q\,) \int \frac{V d^3 k}{(2\pi)^3} \widetilde{\psi}_2^*(\vec k + \vec q\,) \widetilde{\psi}_1(\vec k)\, ,
\end{equation}
where \(\widetilde{\psi}_1\) and \(\widetilde{\psi}_2\) are the momentum-space wavefunctions of the initial and final states, respectively. Here, $V$ is the volume of space, which is defined by $(2\pi)^3\delta^{(3)}(\vec 0)\equiv V$.
Squaring the bound-state amplitude, the following substitution is made:
\begin{equation}
V (2 \pi)^3 \delta^3(\vec k - \vec q - \vec k') |\mathcal M_{\rm free}|^2
\longrightarrow
|\mathcal M_{\rm free}|^2 \times V^2 | f_{1\to 2}(\vec q \,)|^2 \, ,
\end{equation}
where \(f_{1\to 2}(\vec q\,)\) is the atomic form factor:
\begin{equation}
f_{1\to 2}(\vec q \,) = \int \frac{d^3 k}{(2\pi)^3} \widetilde{\psi}_2^*(\vec k + \vec q\,) \widetilde{\psi}_1(\vec k) \, .
\end{equation}
If the detector response is isotropic, as is mostly the case for materials like silicon and germanium, we can replace $f_{1\to 2}(\vec q \,) \to f_{1\to 2}(q \,)$. 
The phase-space integral for the final electron is replaced as follows:
\begin{equation}
\text{free-electron phase space} = V \int \frac{d^3 k'}{(2 \pi)^3} \longrightarrow 1 \, .
\end{equation}
Combining these modifications, the cross section for DM-induced electron excitation is:
\begin{equation}
\sigma v_{1\to 2} = 
\frac{1}{4 E_\chi' E_e'} \int \frac{d^3 q}{(2\pi)^3} \frac{1}{4 E_\chi E_e} 2 \pi \delta(E_i - E_f) \overline{|\mathcal M_{\rm free}(\vec q\,)|^2} | f_{1\to 2}(q \,)|^2 \, .
\end{equation}
Under the non-relativistic approximation, the initial and final energies are:
\begin{equation}
E_i = m_\chi + m_e + \frac{1}{2} m_\chi v^2 + E_{e,1}, \quad
E_f = m_\chi + m_e + \frac{|\vec p - \vec q\,|^2}{2 m_\chi} + E_{e,2} \, .
\end{equation}
We can parametrize the DM-electron coupling with:
\begin{equation}
\label{eq:DMeparam}
\overline{|\mathcal M_{\rm free}(\vec q\,)|^2} \equiv \overline{|\mathcal M_{\rm free}(\alpha m_e)|^2} \times |F_{\rm DM}(q)|^2, \quad
\overline{\sigma}_e \equiv \frac{\mu_{\chi e}^2 \overline{|\mathcal M_{\rm free}(\alpha m_e)|^2}}{16 \pi m_\chi^2 m_e^2} \, .
\end{equation}

For example, let's consider the case of a fermionic DM candidate, $\chi$, which interacts with electrons through a $U(1)_D$ vector mediator $V$.\footnote{The result will be the same for DM that is a complex scalar.} The mediator kinematically mixes with the SM hypercharge $U(1)_Y$ through
\begin{equation}
{\cal L}\supset \frac{\epsilon}{2\cos\theta_W}V_{\mu\nu}F_Y^{\mu\nu}\, ,
\end{equation}
where $\epsilon$ is the kinetic mixing parameter, $\theta_W$ is the Weinberg mixing angle, and $V_{\mu\nu}~(F^{\mu\nu}_Y)$ is the $U(1)_D~(U(1)_Y)$ field strength. 
In this case, 
\begin{align}
\overline\sigma_e = \frac{16\pi\mu^2_{\chi e} \alpha \epsilon^2\alpha_D}{(m_V^2+\alpha^2 m_e^2)^2}
\simeq
\begin{cases}
\frac{16 \pi \mu_{\chi e}^2 \alpha \epsilon^2 \alpha_D}{m_V^4}\,, & m_V \gg \alpha m_e \\
\frac{16 \pi \mu_{\chi e}^2 \alpha \epsilon^2 \alpha_D}{(\alpha \, m_e)^4}\,, & m_V \ll \alpha m_e
\end{cases}\,,
\end{align}
where $\mu_{\chi e}$ is the DM-electron reduced mass and $\alpha_D\equiv g_D^2/4\pi$ with $g_D$ the $U(1)_D$ gauge coupling. 
$F_{\rm DM}(q)$ encodes the $q$-dependence of the DM-electron interaction,
\begin{equation}
\label{eq:vectorFDM}
 F_{DM}(q) = \frac{m_V^2+\alpha^2m_e^2}{m_V^2+q^2} \simeq
\begin{cases}
1\,, & m_V \gg \alpha m_e \\
\frac{\alpha^2 m_e^2}{q^2}\,, & m_V \ll \alpha m_e
\end{cases}\, .
\end{equation}
Note that more generally, $F_{\rm DM}$ can also be a function of DM velocity $v$ \ie $F_{\rm DM}(q, v)$. This can arise from the non-relativistic effective theory of DM operators containing spin, see for example Refs.~\cite{Fan:2010gt,Fitzpatrick:2012ix,Catena:2019gfa,Liang:2024lkk}.

With the substitutions in Eq.~\ref{eq:DMeparam}, the cross section simplifies to:
\begin{equation}
\sigma v_{1\to 2} = 
\frac{\overline{\sigma}_e}{\mu_{\chi e}^2} \int \frac{d^3 q}{4 \pi} \delta \Big(\Delta E_{1\to2} + \frac{q^2}{2 m_\chi} - q v \cos \theta_{q v} \Big) |F_{\rm DM}(q)|^2 | f_{1\to 2}(q \,)|^2 \, .
\end{equation}
The excitation rate is then:
\begin{equation}
R_{1\to2} = \frac{\rho_\chi}{m_\chi} \frac{\overline{\sigma}_e}{8\pi \mu_{\chi e}^2} \int d^3 q \frac{1}{q} \eta\big(v_{\rm min}(q, \Delta E_{1\to2}) \big) |F_{\rm DM}(q)|^2 | f_{1\to 2}(q \,)|^2 \, ,
\end{equation}
where:
\begin{equation}
\eta(v_{\rm min}) = \int \frac{d^3 v}{v} g_\chi(v) \Theta(v - v_{\rm min}), \quad
v_{\rm min}(q, \Delta E_{1\to2}) = \frac{\Delta E_{1\to2}}{q} + \frac{q}{2 m_\chi} \, .
\end{equation}
Note that although $\eta(v_{\rm min})$ is the same quantity as in the previous section on DM-nuclear scattering, the definition of $v_{\rm min}$ has changed. In particular, there is no longer a unique relationship between the energy transfer and the minimum velocity. 

One can generalize the above discussion to account for anisotropic velocity distributions or detector responses, as well as velocity-dependent interactions ($F_{\rm DM}(q,v)\propto q^a v^b$), by making the substitutions~\cite{Lillard:2025aim}
\begin{align}
\eta(v_{\rm min})&\to \eta(\vec q,E) \equiv 2q \int d^3 v v^b g_\chi(\vec v) \delta\left(E+\frac{q^2}{2m_\chi}-\vec q\cdot\vec v\right)\, ,\\
f_{1\to 2}(q) &\to f_{1\to 2}(\vec q)\, ,
\end{align}
where the factor of $2q$ in the first line ensures that the angular average of $\eta(\vec q,E)$ matches the isotropic case. Here, we can define 
\beq
v_{\rm min}(\vec q, E)\equiv \frac{E}{q}+\frac{q}{2m_\chi}+\frac{\vec q\cdot\vec v_E}{q}\, ,
\eeq
where $\vec v_E$ is the Earth's velocity.
Fig.~\ref{fig:DMescattering} shows the status of DM-electron scattering in 2025. 
\begin{figure}
\centering
\includegraphics[width=0.45\linewidth]{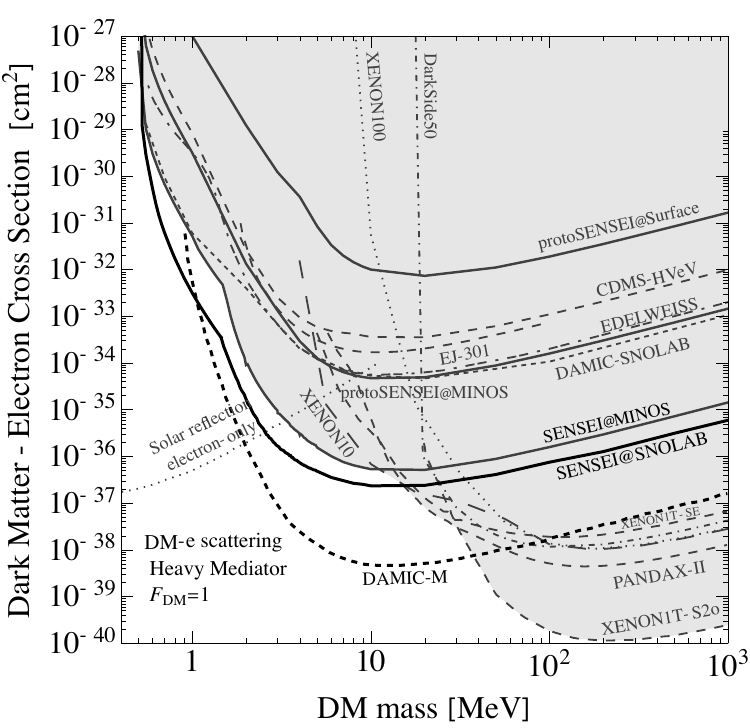}
\includegraphics[width=0.45\linewidth]{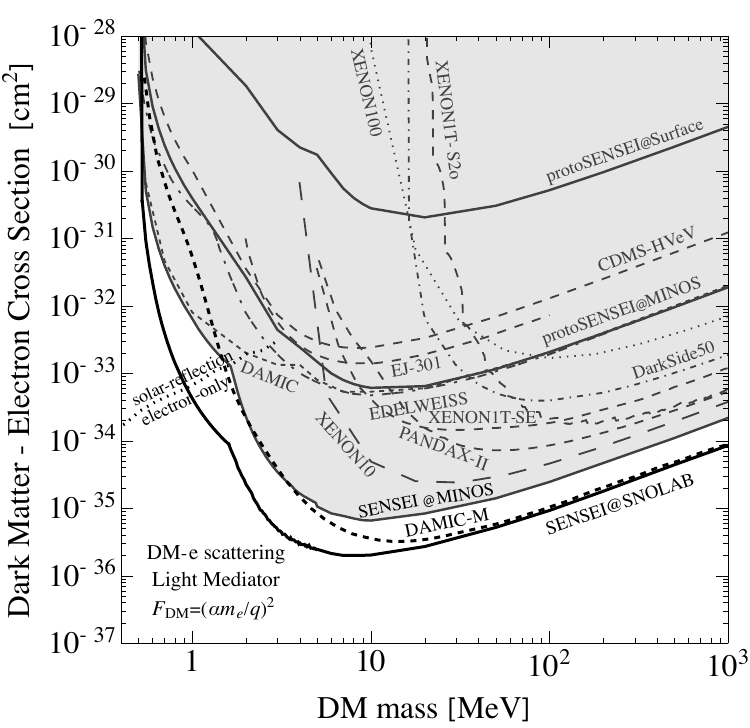}
\includegraphics[width=0.45\linewidth]{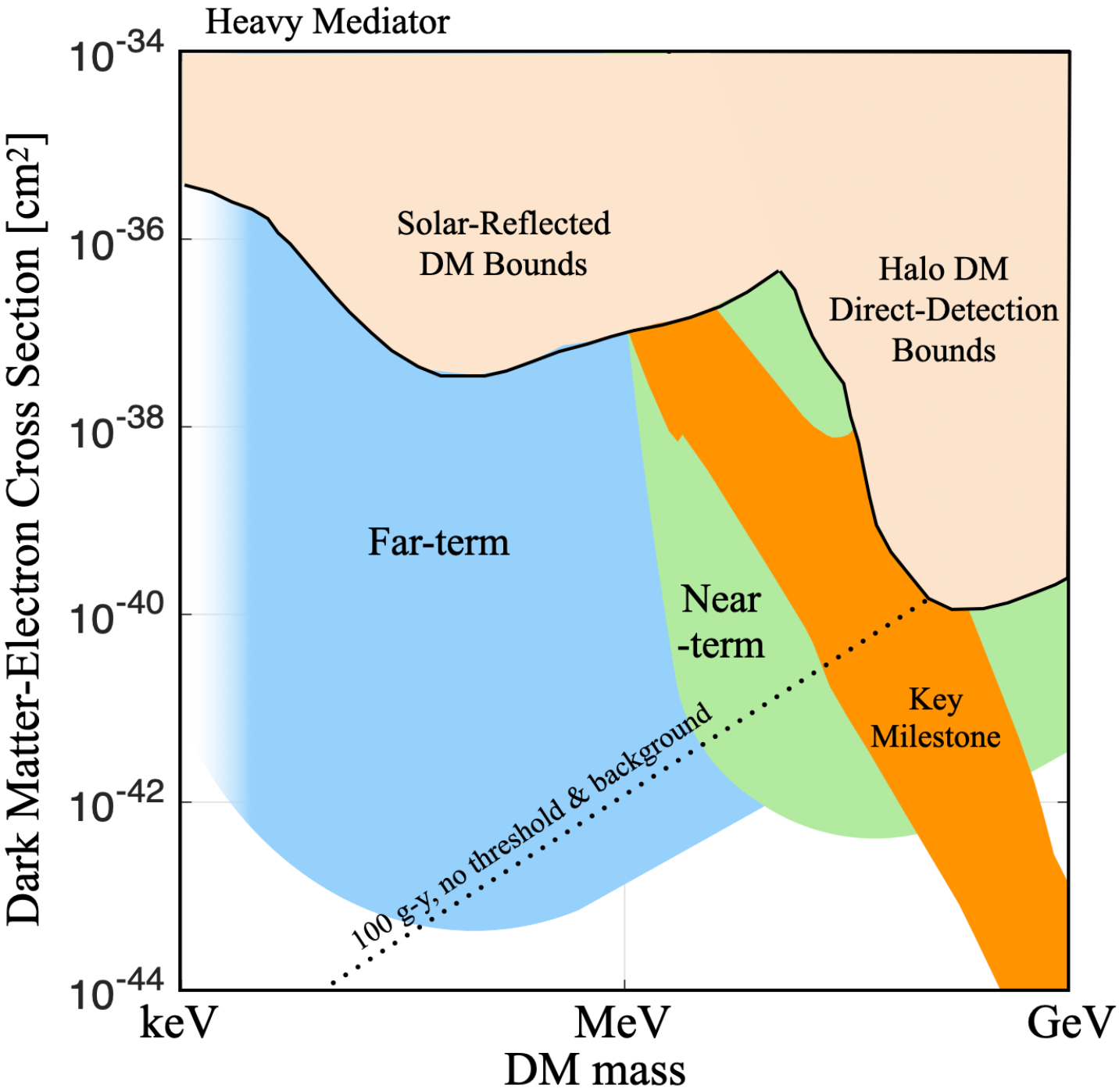}
\includegraphics[width=0.45\linewidth]{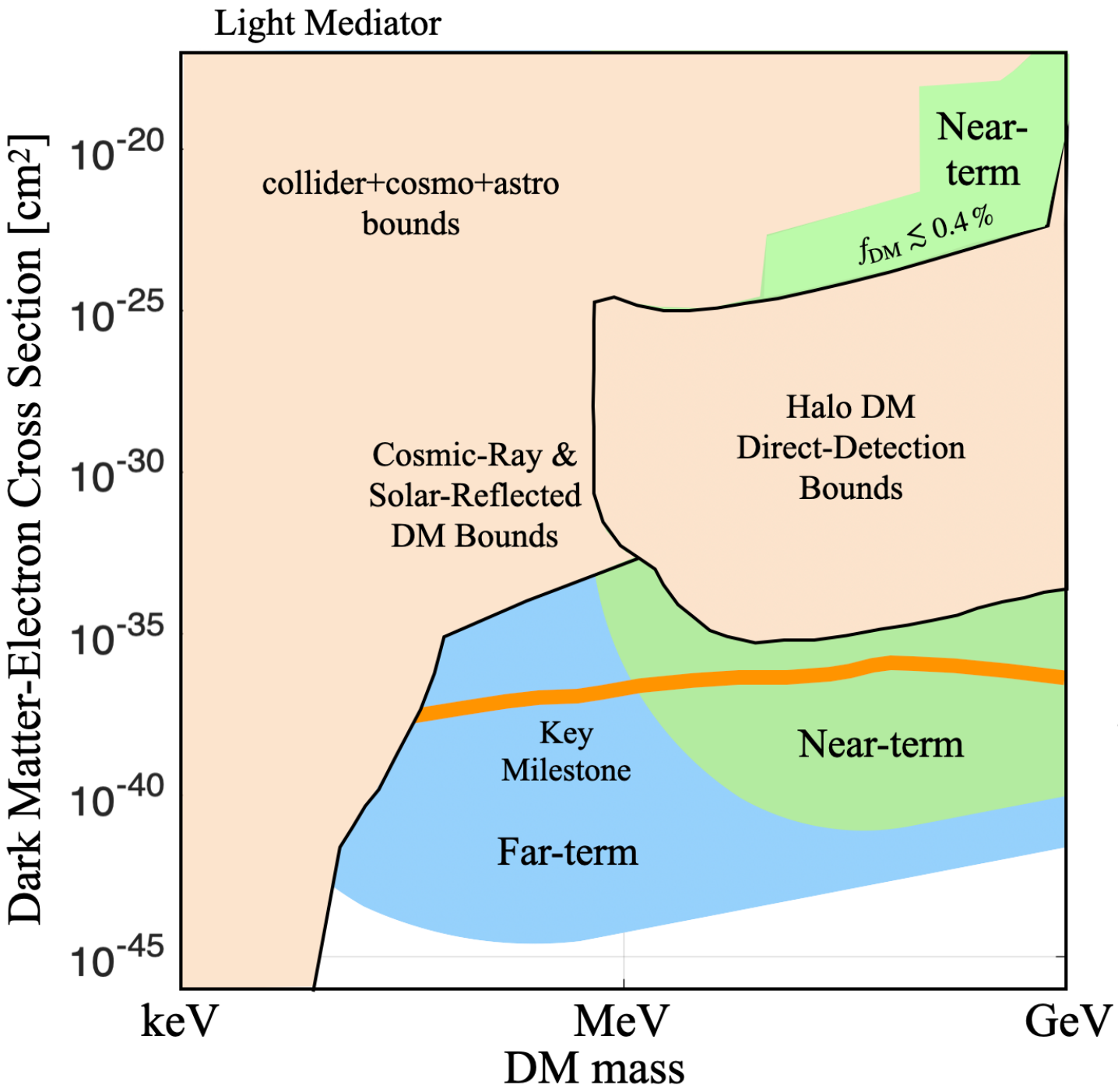}
\caption{Status of DM-electron scattering ({\bf top}) in 2025 and ({\bf bottom}) future sensitivity. The regions labeled ``key milestones" indicate parameter space of specific DM models which gives the correct DM relic abundance. Figures adapted from~\cite{Essig:2022dfa,Cooley:2022ufh}.}
\label{fig:DMescattering}
\end{figure}

 \subsubsection{Bosonic DM absorption}\label{sec:DMabsorption}
We'll conclude this section with an important observation. Experiments designed to detect sub-GeV dark matter (DM) interacting with electrons, such as those sensitive to electron energy depositions, are also capable of detecting another process -- the absorption of bosonic DM by atoms or crystals. 
This highlights a key point: experiments developed for one specific purpose can often be repurposed to explore other phenomena, offering a broader range of potential discoveries.

The absorption of bosonic DM is similar to the photoelectric effect, and its rate can  be derived using the photoabsorption cross-section, $\sigma_{\rm PE}$. 
The photoelectric cross-section can be written in terms of the target material’s electrical response functions~\cite{An:2013yua,Bloch:2016sjj,Hochberg:2016sqx}. Specifically, the absorption rate for photons can be obtained from the polarization tensor of the electromagnetic field, $\Pi$, through the optical theorem:
\begin{equation}
n_e\langle \sigma_{\rm PE} v\rangle=-\frac{{\rm Im}~\Pi(\omega)}{\omega}\, ,
\end{equation}
where $\omega$ is the energy of the absorbed photon. The polarization tensor, $\Pi$, can be written in terms of the complex index of refraction, $\hat n$, which encodes the frequency-dependent response of the target material to an electromagnetic field. The transverse and longitudinal polarization tensors are given by 
\begin{equation}
\Pi_T=\omega^2(1-\hat n^2),~\Pi_L=(\omega^2-|\vec q|^2)(1-\hat n^2)\, ,
\end{equation} 
where $q=(\omega,\vec q)$ is the 4-momentum of the dark photon and $\hat n=1-\frac{r_0}{2\pi}\lambda^2\sum_An_A(f_1^A+if_2^A)$. In the last expression, $n_A$ is the density of atoms of type $A$, $\lambda=2\pi/\omega$ is the wavelength of a photon with energy $\omega$, $r_0=2.82\times 10^{-15}$ m is the classical electron radius, and $f_1, f_2$ are experimentally determined atomic scattering factors. The index of refraction can also be written in terms of the complex conductivity, $\hat\sigma$, through $\hat n^2=1+i\hat\sigma/\omega$. Plugging this into the expression for $\Pi_{T,L}$ and taking the limit where $|\vec q|\ll \omega$, we get the relationship that $\Pi(\omega)\simeq -i\omega\hat\sigma$. Writing out $\hat \sigma=\sigma_1+i\sigma_2$ and substituting in the expression for $\Pi$, we see that 
\begin{equation}
n_e \langle \sigma_{\rm PE} v_{rel}\rangle=\sigma_1\, ,
\end{equation}
Writing the photoelectric cross-section in this way is useful as one can experimentally measure the conductivity of given materials. In this way, one can then relate the rate of DM absorption to experimentally measured quantities.\footnote{This statement is true whenever the DM-electron interactions are sufficiently similar to SM interactions, and has more recently been applied to DM-electron scattering~\cite{Knapen:2021run,Hochberg:2021pkt,Catena:2024rym,Berlin:2025uka}.} 

We can apply this expression to specific examples of bosonic DM. 
Let's first consider the absorption of dark photon ($V$) DM, which interacts with the SM through kinetic mixing,
\begin{equation}
{\cal L}\supset -\frac{\epsilon}{2}F_{\mu\nu}V^{\mu\nu}\, ,
\end{equation}
where $\epsilon$ is the kinetic-mixing parameter and $F_{\mu\nu} (V_{\mu\nu})$ are the field strengths for the photon (dark photon). The absorption cross-section for $V$ can be expressed in terms of the photoelectric cross-section $\sigma_{\rm PE}$:
\begin{equation}
\sigma_{V}(E_{V} = m_{V})\,v_{V} \simeq \epsilon^2 \sigma_{\rm PE}(E = m_{V}) \,,
\end{equation}
where $v_{V}$ is the velocity of the dark photon. 

For dark photons, one needs to take care to include in-medium effects, which can substantially alter the dark photon polarization tensor. These effects are especially pronounced for particles produced in extreme environments, such as dark photons produced in the Sun, but are also relevant for non-relativistic DM. These in-medium effects give rise to an {\it effective} mixing angle,
\begin{equation}
\epsilon_{\rm eff_{T,L}}^2=\frac{\epsilon^2m_V^4}{[m_V^2-{\rm Re}~\Pi_{T,L}(m_V)]^2+[{\rm Im}~\Pi_{T,L}(m_V)]^2}\, ,
\end{equation}
and so the corresponding absorption rate per atom is:

\begin{align}
{\rm Dark~photon~absorption~rate~per~atom} \simeq n_V \times \epsilon_{\rm eff}^2\sigma_1\,,
\end{align}
where $n_V$ is the number density of dark photons.

Pseudoscalar particles, such as axions and axion-like particles, can undergo a similar ``axioelectric" effect in which they are similarly absorbed by bound electrons. Given the effective Lagrangian
\begin{equation}
{\cal L}_a=\frac{1}{2}\partial_\mu a\partial^\mu a-\frac{1}{2}m_a^2 a^2+i g_{aee}a\bar e\gamma_5e\, ,
\end{equation}
the absorption cross-section for this scenario is given by
\begin{equation}
\sigma_a(E)v_a\simeq \frac{3}{4}\frac{g_{aee}^2}{4\pi\alpha_{\rm EM}}\frac{E^2}{m_e^2}\left(1-\frac{1}{3}v_a^{3/2}\right)\sigma_{\rm PE}(E)\, ,
\end{equation}
where $E$ and $v_a$ are the energy and velocity of the axion, respectively. This leads to the following expression for the absorption rate of axions or axion-like particles,
\begin{equation}
{\rm Axion~absorption~rate~per~atom} \simeq n_a \frac{3m_a^2}{4m_e^2}\frac{g_{aee}^2}{4\pi\alpha_{\rm EM}}\sigma_1\,,
\end{equation}
where $n_a$ is the number density of axions.
Fig.~\ref{fig:DMabsorption} shows the status of dark photon DM bosonic absorption in 2025. 

Current and upcoming direct detection experiments employ a variety of target materials (noble liquids, semiconductors, organic molecules, superfluids) and detection techniques (ionization, scintillation, phonons) to maximize sensitivity across different DM mass ranges. An in-depth discussion of DM direct detection searches using condensed matter systems can be found in Ref.~\cite{Kahn:2021ttr}.
With continued improvements in detector technology and background mitigation, direct detection remains one of the most promising avenues for discovering the elusive nature of DM. 
\begin{figure}
\centering
\includegraphics[width=0.45\linewidth]{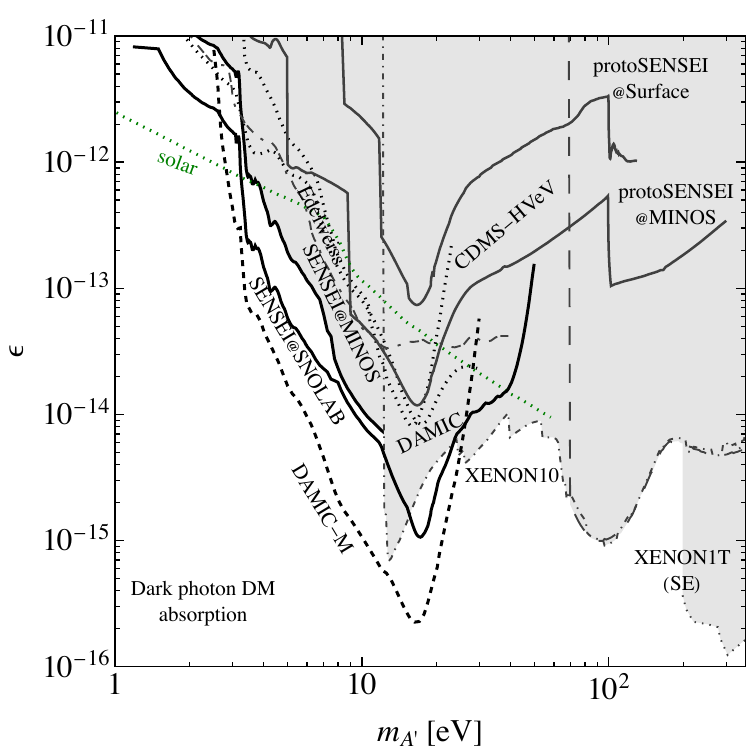}
\includegraphics[width=0.45\linewidth]{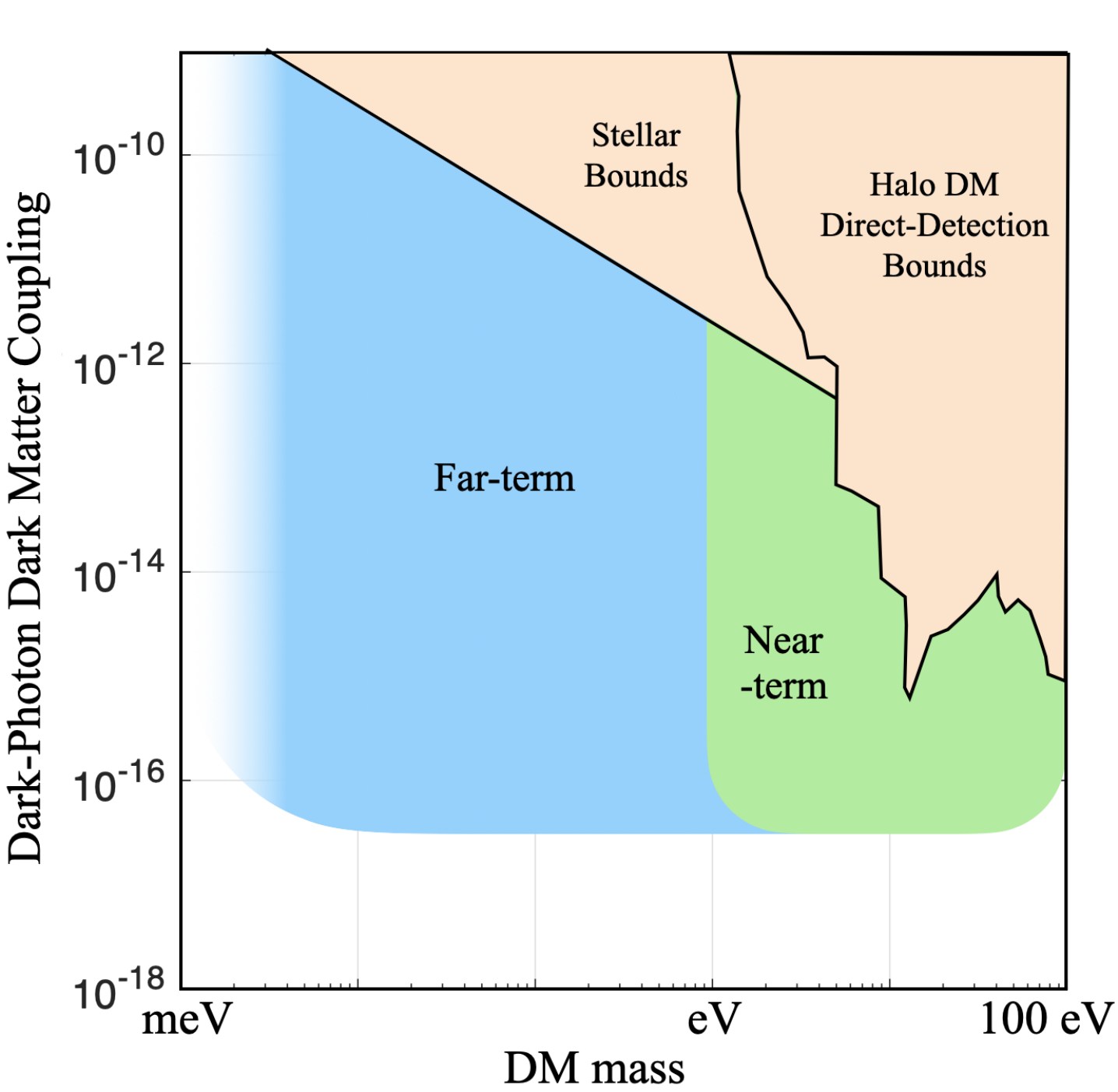}
\caption{Dark photon dark matter absorption ({\bf left}) status as of 2025 and ({\bf right}) future sensitivity. Figures adapted from~\cite{Essig:2022dfa,Cooley:2022ufh}}
\label{fig:DMabsorption}
\end{figure}

\section*{Acknowledgments}
I would like to express my gratitude to the scientific organizers of the 2024 TASI, Nathaniel Craig, Tongyan Lin, and Jesse Thaler, as well as to the TASI directors, Tom DeGrand, Oliver DeWolfe, and Ethan Neil, for the opportunity to deliver these lectures. I am also grateful to Francesco D’Eramo for his kind hospitality and for arranging my lecture visit at the Università degli Studi di Padova. I thank the students for their enthusiastic engagement, which made the lecture series especially enjoyable. I am particularly indebted to Pouya Asadi, Ben Lillard, Thomas Schwemberger, and Edoardo Vitagliano for their insightful feedback on a working draft. I acknowledge the CERN TH Department for hospitality where this work was completed. This work was supported in part by NSF CAREER grant PHY-1944826.
\bibliographystyle{jhep}
\bibliography{refs}
\end{document}